\documentclass[]{aa}

\usepackage{graphicx}
\usepackage{txfonts}
\usepackage{color}

\begin{document}
 
\title{New evolutionary sequences for extremely low mass white dwarfs:} 
\subtitle{Homogeneous mass and age determinations, and asteroseismic prospects}
\author{Leandro G. Althaus,
        Marcelo M. Miller Bertolami \and
        Alejandro H. C\'orsico}
\institute{Grupo de Evoluci\'on Estelar y Pulsaciones. Facultad de Ciencias Astron\'omicas y 
           Geof\'{\i}sicas,
           Universidad Nacional de La Plata, CONICET-CCT.
           Paseo del Bosque s/n,
           1900 La Plata, Argentina\\
           \email{althaus,mmiller,acorsico@fcaglp.unlp.edu.ar}}          
\date{\today}

\abstract{ The number of detected extremely low mass (ELM) white dwarf
  stars has increased drastically in recent year thanks to the
  results of many surveys. In addition, some of these stars have been found
  to exhibit pulsations, making them potential targets for
  asteroseismology.}  {We provide a fine and homogeneous grid of
  evolutionary sequences for helium (He) core white dwarfs for the
  whole range of their expected masses ($0.15\lesssim
  M_*/M_{\sun}\lesssim 0.45$), including the mass range for ELM white dwarfs ($
  M_*/M_{\sun}\lesssim 0.20$). The grid is appropriate for mass and age
  determination of these stars, as well as to study their adiatabic
  pulsational properties.}  {White dwarf sequences have been computed
  by performing full evolutionary calculations that consider the main
  energy sources and processes of chemical abundance changes during
  white dwarf evolution. Realistic initial models  for the evolving white
  dwarfs have been obtained by computing  the non-conservative evolution of
 a binary system consisting of an initially $1 M_\odot$ ZAMS star and a  $1.4 M_\odot$
 neutron star for various initial orbital periods. To derive cooling ages and masses for He-core
  white dwarf we perform a least square fitting of the $M(T_{\rm eff}, g)$
  and ${\rm Age}(T_{\rm eff}, g)$ relations provided by our sequences by using 
  a scheme that takes  into account the time spent by models in different regions
  of the $T_{\rm eff}-g$ plane. This is particularly useful  when multiple solutions for
cooling age and mass determinations are possible in the case of CNO-flashing sequences.
 We also explore in a preliminary way 
  the adiabatic pulsational properties of models near the critical mass for the
  development of CNO flashes ($\sim 0.2 M_\odot$). This is motivated by the discovery  
of pulsating   white dwarfs with stellar masses near this threshold value. } {We obtain 
reliable and 
  homogeneous
  mass and cooling age determinations for 58 very low-mass white dwarfs, including 3
  pulsating stars.  Also, we  find substantial 
  differences in the period spacing distributions of $g$-modes 
  for models with stellar masses near $\sim 0.2 M_\odot$, which could be 
  used as a seismic tool to distinguish stars that have undergone 
  CNO flashes in their early cooling phase from those that have not.
  Finally, for an easy application of our results, we provide
  a reduced  grid of values useful to obtain masses and ages of He-core white
  dwarf.}  
  {}
\titlerunning{New evolutionary sequences for ELM white dwarfs.}
\authorrunning{Althaus et al.}  \maketitle


\section{Introduction}
\label{intro}

White dwarf  stars are  routinely used to  constrain the age  and past
history of the Galactic populations, including the solar neighborhood,
open and globular clusters (Von  Hippel \& Gilmore 2000; Hansen et al.
2007;  Winget  et  al.   2009;  Garc\'\i  a-Berro  et  al.   2010; Bono et al.
2013,  and
references therein).  In addition,  they are used to place constraints
on  properties  of  elementary  particles  such as  axions  (Isern  et
al.  1992;  Isern  et  al.   2008; C\'orsico  et  al.   2012a,b),  and
neutrinos  (Winget  et  al.   2004)  or  on  alternative  theories  of
gravitation (Garc\'\i a--Berro et  al.  1995; Garc\'\i a--Berro et al.
2011).  These and other potential applications of white dwarfs require
a detailed and  precise knowledge of the main  physical processes that
control their evolution (see Fontaine \& Brassard 2008; Winget \& Kepler 2008 and 
Althaus et al. 2010a for a review).

The white  dwarf mass distribution includes  a  population of low-mass
remnants, most of them expected to have a helium (He) core in their interiors (see Kepler
et al.  2007).  In  recent years, the  number of  detected white dwarfs  
with very  low stellar masses,  commonly referred  to as
Extremely Low Mass (ELM) white dwarfs, has increased considerably thanks
to the result of many surveys, in particular the  ELM survey,  
and the  SPY and  WASP surveys  (see Koester  et al.
2009;  Brown et  al. 2010;  Maxted et  al. 2011,  Brown et  al. 2012).
Because of  the very low  stellar mass values that  characterize these
ELM  white dwarfs (lower  than about  0.20 $M_{\sun}$\footnote{This corresponds
approximately to the value of the mass threshold for the occurrence of 
CNO  flashes on the cooling branch, see later in the paper.}),  they are
believed to  be the result  of compact binary evolution,  during which
the envelope  of a red giant  star is removed before  the core reaches
enough mass to  ignite helium. The evolution of  He-core white dwarfs
has been studied by Driebe et al. (1998), Sarna et al. (2000), Althaus et al. (2001), 
Serenelli et al.  (2002), Nelson et al. (2004), Benvenuto \& De Vito (2005), 
Panei et al. (2007), and more recently by Gautschy (2013).

A major step towards the  understanding of the formation and evolution
of ELM white  dwarfs has been the recent  discovery of three pulsating
He-core  white  dwarfs  with  stellar  masses below  0.23  $M_{\sun}$  and
effective  temperatures  less  than  10,000K  (Hermes  et  al.   2012,
2013). This new class of  pulsating white dwarfs most probably belongs
to  a low-mass  extention of  the ZZ  Ceti instability  strip  to much
cooler effective temperatures. It  is expected that the application of
the tools of asteroseismology to  these and other pulsating ELM white
dwarfs   being found in the future will  reveal   details  of   their  internal   structure  and
evolutionary status; see Steinfadt  (2010) and C\'orsico et al. (2012)
for  the  first  exploration  of  the adiabatic  properties  of  these
objects. 

The  development  of  a  fine  and homogeneous  grid  of  evolutionary
sequences  for  He-core white  dwarfs,  and  in  particular ELM  white
dwarfs,  derived consistently  from  binary evolution  is therefore  a
pressing  necessity  for  precise  mass and  cooling age  determinations,  and
asteroseismological inferences for these  stars. This is precisely the
main aim  of this  paper. As shown  by Sarna  et al.  (2000)  a proper
treatment of  the binary evolution  leading to the formation  of these
stars  is of  utmost importance  for the  correct assessment  of their
evolutionay properties.  This is  particularly true regarding the mass
of  the  hydrogen  envelope that  is  left  after  the end  of  binary
evolution.  In particular, Sarna et  al. (2000) found that for stellar
masses   lower  than   $\approx  0.25   M_{\sun}$,   binary  evolution
calculations yield more massive hydrogen envelopes than those found by
arbitrarily abstracting mass to a  red giant star.  These authors also
reported   that  during   the  evolution   through  the   stages  at
approximately  constant  luminosity  that  follow the  end  of  binary
evolution,  and also  during most  of  the cooling  branch, an  active
hydrogen burning  shell remains, delaying  the evolution of  ELM white
dwarfs by several Gyr.  In this connection, Sarna et al.  (2000) found
that the  evolutionary times during the stages  of constant luminosity
are strongly dependent on the mass of the white dwarf remnant.  Hence,
for  a proper  assessment of  evolutionary timescales  and mass-radius
relations, the binary nature that  leads to the formation of ELM white
dwarfs  must be  taken into  account. This  has been  the  approach we
adopted in this  work to generate the initial  models for the evolving
white   dwarfs.      Specifically,   we   have   considered   the
  non-conservative  angular-momentum  evolution of a binary system consisting 
of  an  initially  $1
  M_\odot$  ZAMS  component and  a  $1.4
  M_\odot$ neutron  star as the other component.   In particular, our
white dwarf sequences start shortly after the end of Roche lobe phase.

In addition to the discussion  of the evolutionary expectations of our
sequences,  we  extend  the  scope   of  the  paper  by  presenting  a
preliminary exploration of the adiabatic pulsational properties of our
evolutionary models near the critical stellar mass for the development
of CNO  flashes  on the cooling branch. We  show in particular that
seismic  tools exploiting  these  properties can  be  used to  extract
information about the occurrence of CNO flashes in prior stages, and 
the consequent age dichotomy expected in low-mass He-core white dwarfs.
Our interest in this aspect is motivated by the fact that the three pulsating ELM white
dwarfs discovered by Hermes et al. (2013) have stellar mass precisely near
the threshold value for the occurrence of CNO flashes. 
 
The paper is outlined as follows. In  Sect. \ref{tools} we briefly describe 
the input physics of the evolutionary code employed in our
calculations, and the procedure and assumptions we consider for the
computation of binary evolution that leads to the formation of ELM
white dwarfs. In total, 9 initial ELM white dwarf models with stellar
masses between 0.155 and 0.20 $\, M_{\sun}$ have been derived for
initial orbital periods at the beginning of the Roche lobe phase in
the range 0.9 to 2 d.  Additional binary calculations have been  conducted
for binary configurations with initial orbital periods up to 300d to cover the
whole mass range expected for He-core white dwarfs. In
Sect.  \ref{results} the main evolutionary characteristics of our
sequences are presented, while in Sect. \ref{MassAge} we describe the
method adopted to derive from these sequences ages and masses for He-core 
white dwarfs. In this
Section we also present a new and homogeneous determination of masses and,
for the first time, cooling ages of a large sample of recently discovered 
low-mass (most of them ELM) white dwarfs, and a simple
algorithm for an easy use of our sequences as well. In Sect. 
\ref{pulsation} we describe the main basic
expectations of our sequences for the adiabatic pulsation properties
of models representative of the observed pulsating ELM white dwarfs.

\begin{table}[h]
\caption{\label{tab1} Characteristics of our initial He-core white dwarf models.}
\centering
\begin{tabular}{lccccc}
\hline\hline
$M_{\rm f} (M_{\sun})$ &$P_{\rm i}$(d)& $P_{\rm f}$(d) & $X_{\rm f}^{\rm surf}$
& $M_{\rm H}$ & H flash\\
\hline
0.15540 & 0.90 & 0.23 & 0.365 & $4.34 \times 10^{-3}$ & No\\
0.16115 & 0.95 & 0.32 & 0.376 & $4.19 \times 10^{-3}$ &No\\
0.16499 & 1.0  & 0.35 & 0.390 & $4.09 \times 10^{-3}$ &No\\
0.17064 & 1.1  & 0.44 & 0.405 & $3.94 \times 10^{-3}$ &No\\
0.17624 & 1.2  & 0.54 & 0.423 & $3.80 \times 10^{-3}$ &No\\
0.18213  & 1.3  & 0.80 & 0.440 & $3.66 \times 10^{-3}$&Yes\\
(0.18050) &   &  &  & & \\
0.18685  & 1.4  & 1.14 & 0.453 & $3.55 \times 10^{-3}$& Yes\\
(0.18630) &   &  &  & & \\
0.19210  & 1.55 & 1.54 & 0.466 & $3.42 \times 10^{-3}$& Yes\\
(0.19173) &   &  &  & & \\
0.20258  & 2    & 2.47 & 0.490 & $3.22 \times 10^{-3}$&Yes \\
(0.20187) &   &  &  & & \\
0.23903  & 5.2    & 10.32 & 0.701 & $2.86 \times 10^{-3}$&Yes \\
(0.23887) &   &  &  & & \\
0.27242  & 10   & 20.4 & 0.715 & $2.02 \times 10^{-3}$& Yes\\
(0.27065) &   &  &  & & \\
0.32079  & 40   & 78.7 & 0.715 & $1.18 \times 10^{-3}$& Yes\\
(0.32048) &   &  &  & & \\
0.36304  & 100  & 187.7& 0.715 & $7.85 \times 10^{-4}$&Yes\\
(0.36242) &   &  &  & & \\
0.43520 & 300 & 520   & 0.715 & $4.33 \times 10^{-4}$& No\\
\hline
\end{tabular}
\tablefoot{\\
$M_{\rm f} (M_{\sun})$:Secondary mass at the end of Roche lobe 
overflow.\\
$P_{\rm i}$: Initial orbital period of the system (in days). \\
$P_{\rm f}$: Final orbital period at the end of Roche lobe overflow (in days). \\
$X_{\rm f}^{\rm surf}$: Hydrogen surface abundance at the end of
Roche lobe overflow.\\
$M_{\rm H}$: Mass (solar masses) of the hydrogen content at the point
of maximum effective temperature at the beginning of the first
cooling branch.\\
H Flash: Occurrence of CNO flashes on the early white dwarf
cooling branch.
}
\label{tabla}
\end{table}


\section{Numerical tools}
\label{tools}

The evolutionary  calculations presented in  this work have  been done
using  the {\tt  LPCODE}  stellar evolutionary  code  (Althaus et  al.
2003, 2005, 2012).   This code has recently been  used to produce  very
accurate  white dwarf  models  --  see
Garc\'\i a--Berro et  al.  (2010), Althaus et al.   (2010b), Renedo et
al.   (2010), Miller Bertolami  et al.   (2011ab),  and references
therein.  The code  has  also been  used  to study  the formation  of
subdwarf  stars from  the post-red  giant branch  hot-flasher scenario
(Miller Bertolami et  al.  2008), and the role  of thermohaline mixing
for the  surface composition  of low-mass giant  stars (Wachlin  et al.
2011).  A complete  description  of the  input  physics and  numerical
procedures can be  found in these works.  The nuclear network
accounts  explicitly  for the  following  elements: $^{1}$H,  $^{2}$H,
$^{3}$He, $^{4}$He, $^{7}$Li,  $^{7}$Be, $^{12}$C, $^{13}$C, $^{14}$N,
$^{15}$N,  $^{16}$O,  $^{17}$O,   $^{18}$O,  $^{19}$F,  $^{20}$Ne  and
$^{22}$Ne,  together  with  34  thermonuclear reaction  rates for  the
pp-chains,  CNO bi-cycle,  helium  burning, and  C  ignition that  are
identical  to those  described in  Althaus  et al.   (2005), with  the
exception  of  $^{12}$C$\  +\  $p$  \rightarrow \  ^{13}$N  +  $\gamma
\rightarrow    \    ^{13}$C    +    e$^+   +    \nu_{\rm    e}$    and
$^{13}$C(p,$\gamma)^{14}$N,  which  are  taken  from  Angulo  et  al.
(1999).  Radiative opacities  are those  of OPAL  (Iglesias  \& Rogers
1996).   Conductive
opacities are  from Cassisi et al.  (2007). The equation
of  state during  the  main sequence  evolution  is that  of OPAL  for
hydrogen- and helium-rich composition, and a given metallicity.
Finally, updated  low-temperature molecular opacities  with varying  C/O ratios
are used.  To this end,  we have adopted the low temperature opacities
computed  at  Wichita State  University  (Ferguson  et  al. 2005)  and
presented  by Weiss \&  Ferguson (2009). In
{\tt LPCODE} molecular opacities are computed by adopting  the opacity tables
with the correct abundances of the unenhanced metals (e.g. Fe) and C/O
ratio.  Interpolation is carried  out by  means of  separate cuadratic
interpolations in  $R=\rho/{T_6}^3$, $T$ and $X_{\rm  H}$, but linearly
in $N_{\rm C}/N_{\rm O}$.

For the  evolutionary stages following the  end of mass  loss, and for
the  white dwarf  regime, we  use the  equation of  state of  Magni \&
Mazzitelli (1979) for  the whole star.  We also  take into account the
effects of  element diffusion due to  gravitational settling, chemical
and  thermal  diffusion  of  $^1$H,  $^3$He,  $^4$He,  $^{12}$C,
  $^{13}$C,  $^{14}$N and  $^{16}$O, see  Althaus et  al.   (2003) for
  details.   In  particular,  the  metal  mass fraction  $Z$  in  the
envelope  of  our models  is  specified by  scaling  it  to the  local
abundance of the CNO elements at  each layer.  For the white dwarf 
regime and for effective temperatures lower than 10,000 K, outer boundary
conditions  for the evolving models  are   derived    from    non-grey   model    
atmospheres (Rohrmann et al. 2012). Recently, {\tt LPCODE} has been tested
against other white dwarf code. Uncertainties in white dwarf cooling ages
arising from different numerical implementations of stellar evolution equations
were found to be below 2 per cent (Salaris et al. 2013).

Realistic initial models of ELM  white dwarfs have been obtained by
mimicking the binary evolution of progenitor stars. Since
hydrogen shell burning is the main source of star luminosity
during most of the evolution of ELM white dwarfs, the computation of
realistic initial white dwarf structures is a fundamental issue, in
particular concerning the correct assessment of the hydrogen envelope
mass left by progenitor evolution (see Sarna et al. 2000).  We assume
that the evolution of the binary system is fully non-conservative,
i.e., the total mass and angular momentum of the system are not
conserved. It  is  worth  mentioning  that  changes  in  the  orbital  separation
resulting from changes in the mass  assumed to be lost from the system
are small (see Sarna et al.  2000 for details).
To this end, we follow the formalism of Sarna et
al. (2000). We will denote with $M_1$ the mass of the secondary
(mass-losing) star, $M_2$ the mass of the neutron star (primary), and
with $\dot M_1$ the mass-loss rate of the secondary. The change of the
total orbital angular momentum ($J$) of the binary system can be
written

\begin{figure}
\centering
\includegraphics[clip,width=8.5 cm]{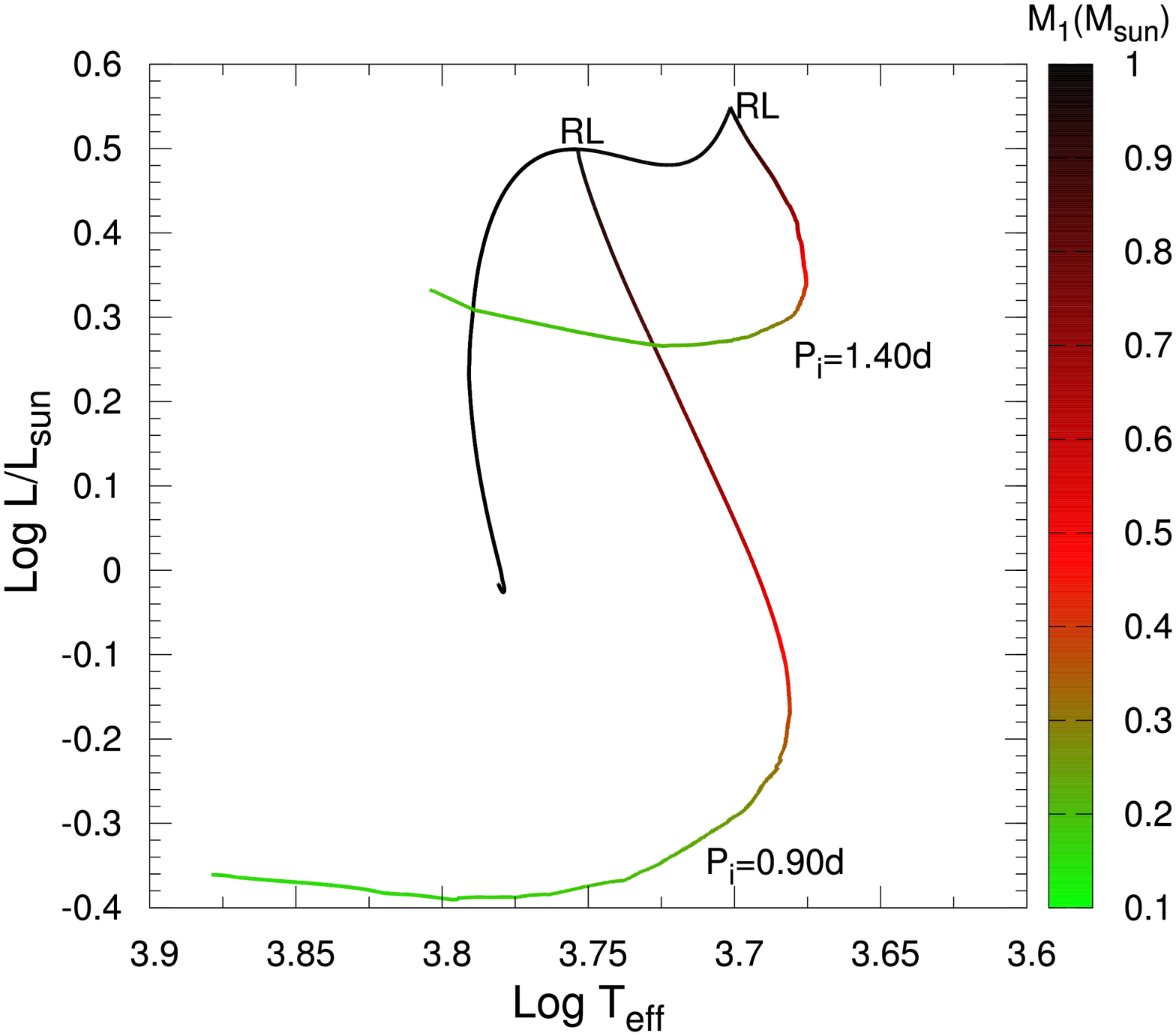}
\caption{Hertzsprung-Russell diagram for the evolution of the
initially 1.0 $M_{\sun}$ secondary star. The tracks corresponding to
initial orbital periods of 0.90 and 1.40d are depicted. The final mass
at the end of Roche lobe overflow is, respectively, 0.15540 and 0.18685 
$M_{\sun}$. RL marks the locations in the diagram when the secondary star
fills its Roche lobe for the first time and mass loss begins. The color scale to 
the right shows the stellar mass of the 
secondary star.}
\label{prewd}
\end{figure}

\begin{equation}
\frac{\dot  J}{J}=  \frac{\dot J_{ML}}{J}  +  \frac{\dot J_{GR}}{J}  +
\frac{\dot J_{MB}}{J}\, ,
\end{equation}

where   $\dot J_{ML}$, $\dot J_{GR}$, and  $\dot J_{MB}$, are respectively,  
the angular momemtum loss from the system due to mass
loss, gravitational wave radiation and magnetic braking (which is relevant 
when the secondary has  an outer convection zone). To compute these quantities
we follow Sarna et al. (2000), see also Muslimov \& Sarna (1993)

\begin{equation}
\frac{\dot J_{ML}}{J} = \frac{M_2}{M_1 (M_1+M_2)} {\dot M_1}\, {\rm yr^{-1}}.
\end{equation}
\begin{equation}
\frac{\dot J_{GR}}{J}  = - 8.5  \times 10^{-10} \frac{M_1  M_2 (M_1+M_2)}{a^4}\,
{\rm yr^{-1}},
\end{equation}

\begin{equation}
\frac{\dot J_{MB}}{J}  = - 3  \times 10^{-7} \frac{(M_1+M_2)^2  R^4_1}{M_1 M_2
a^5}\, {\rm yr^{-1}}\, ,
\end{equation}

where $a$  is the semiaxis  of the orbit and $R_1$ the radius of the secondary.
All quantities are given in solar units.  The mass-loss rate from the secondary
is  calculated   as  in   Chen  \&  Han   (2002).   Mass loss is 
considered as long as the secondary fills its Roche lobe
$r_L$, given by

\begin{equation}
r_L = a \frac{0.49 q^{2/3}}{0.6 q^{2/3} + \rm{ln}(1+q^{1/3})},
\label{eq.egg}
\end{equation}
where $q=M_1/M_2$  is the mass ratio. The semiaxis of the orbit is found
by integrating the equation for  the rate of change of $a$, which in the 
case that the  mass   lost  by  the  secondary  is
completely lost from the system, i.e. nothing of the mass lost by
the secondary is accreted by the primary, is given by (see Muslimov 
\& Sarna 1993)

\begin{equation}
\frac{1}{2}\frac{\dot a}{a}=  \frac{\dot J}{J}  - \left(\frac{1}{M_1} -  \frac{
1}{2(M_1+M_2)}\right) \dot M_1
\end{equation}

Mass loss is continued until the secondary star shrinks  within 
its Roche lobe. It may happen
that as  a result of a  hydrogen thermonuclear flash  occurring on the
white dwarf  cooling branch, the  secondary fills for the  second time
its  Roche lobe  during the  flash.  This  being the  case,  mass loss
becomes operative again on a very short timescale.  We want to mention
that  the   present  treatment  for   mass  loss  is   not  completely
self-consistent in the sense that the mass-loss rate is not considered
as  a unknown  quantity during  the iteration  procedure to  solve the
stellar structure equations, but  instead is fixed beforehand for each
model.   Nonetheless,  it  constitutes  a better  approach  to  derive
physically  sound  initial ELM  white  dwarf  models than  arbitrarily
removing mass to a evolving low-mass star. It is enough for the purpose
of  this  work,  the aim  of  which  is  focused  on the  cooling  and
structural properties of ELM white dwarfs.

All of our  He-core white dwarf initial models have been derived
from evolutionary calculations for binary systems consisting of an
evolving low-mass component of initially $1\, M_{\sun}$ and a $1.4\, M_{\sun}$ 
neutron star as the other component. Metallicity
is assumed to be Z=0.01. A
total of 14 initial He-core white dwarf models with stellar
masses between 0.155 and  0.435 $M_{\sun}$ have been derived for
initial orbital periods at the beginning of the Roche lobe phase in
the range 0.9 to 300 d. In particular, 9 sequences span the range
  of masses corresponding to ELM-white dwarfs ($M\lesssim 0.20 M_{\sun}$).
At this point, it is worth mentioning that the envelope mass of
the resulting white dwarf, a key factor in dictating the cooling times, is only
weakly dependent on the initial mass of the secondary (mass-losing) star (Nelson
et al. 2004). However, different angular-momentum loss prescriptions due to mass loss, 
which could have
an impact on the final envelope mass, have not been explored in this paper. 

\begin{figure*}
\centering
\includegraphics[clip,width=17cm]{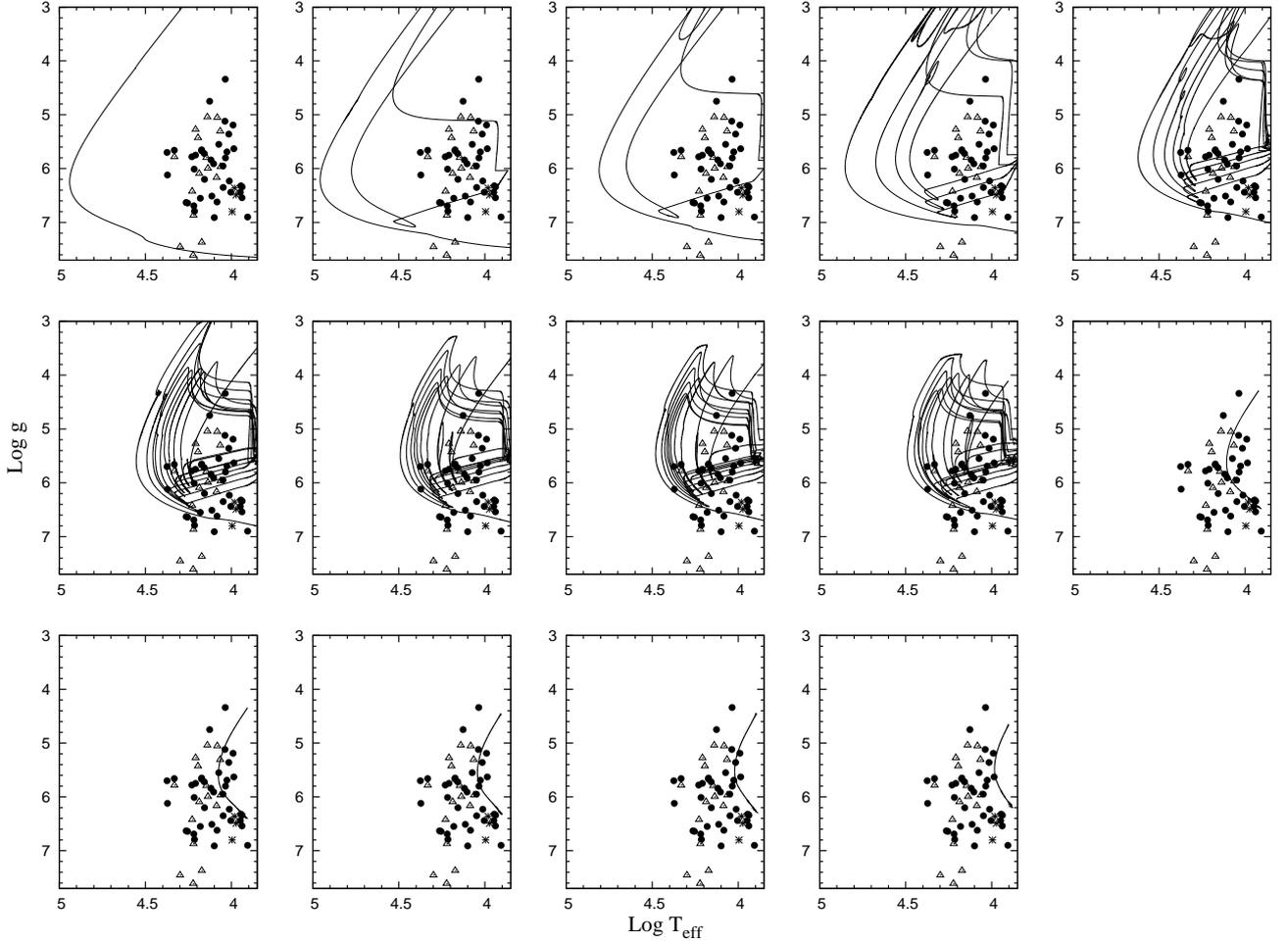}
\caption{Surface gravity - effective temperature diagrams for our
  He-core white dwarf sequences. From left to right and from top to
  bottom, diagrams correspond to sequences of masses: 0.43523,
  0.36304, 0.32079, 0.27242, 0.23903, 0.20258, 0.19210, 0.18685,
  0.18213, 0.17624, 0.17064, 0.16499, 0.16115,
  0.15540$M_{\sun}$. Sequences with masses in the range
  $0.18M_{\sun}\lesssim M_{\rm WD}\lesssim 0.4 M_{\sun}$ undergo
  CNO flashes during the early cooling phase leading to the apparent
  loops in the $g$-$T_{\rm eff}$ diagram. Filled circles and triangles correspond, respectively, 
to the observed post-RGB low-mass stars from Silvotti et al. (2012) and  Brown et al. (2013), 
and asteriks to the pulsating ELM white dwarfs detected by  Hermes et al. (2013), see
Table \ref{Tab:ELMs}.}
\label{Fig:g-teff-ELMs}
\end{figure*}

In  Table \ref{tabla},  we provide  some main  characteristics  of our
whole   set  of   initial   He-core  white   dwarf   models  we   have
calculated.  From  left to  right  we list  the  stellar  mass of  the
secondary  star  at  the  end  of  Roche  lobe  overflow  ($M_{\rm  f}
(M_{\sun})$); the  initial orbital period  (in days) of the  system at
the  beginning of  the mass  transfer phase  ($P_{\rm i}$);  the final
orbital period  (in days) at the  end of Roche  lobe overflow ($P_{\rm
  f}$); the  surface hydrogen  abundance (by mass)  at the end  of the
Roche lobe overflow ($X_{\rm f}^{\rm surf}$), and the mass of the total
hydrogen content at the point  of maximum effective temperature at the
beginning  of the  first cooling  branch ($M_{\rm  H}$).   The further
evolution of these initial models  has been computed down to the range
of  luminosities  of  cool  white  dwarfs,  including  the  stages  of
multiple thermonuclear CNO flashes during the  beginning of cooling
branch.  The numbers  between brackets in the first  column denote the
stellar mass  of the remnant  that is left  after the occurrence  of a
second stage of  Roche lobe overflow during the  CNO flash.
Indeed, during the course of the CNO flash, the white dwarf
remnant may be forced to fill  its Roche lobe again, thus leading to a
new mass transfer  episode. Note also from Table  \ref{tabla} that, in
good agreement  with previous studies  (Sarna et al. 2000;  Althaus et
al. 2001) there exists a threshold in the stellar mass value (at $\sim
0.18 M_{\sun}$), below which CNO flashes are not expected.

\section{Evolutionary results}
\label{results}

As an  example of the evolution  of the white  dwarf progenitor during
the   mass  transfer   stage,  we   show  in   Fig.   \ref{prewd}  the
Hertzsprung-Russell  diagram for  the evolution  of the  initially 1.0
$M_{\sun}$ secondary star. Evolution  of the secondary starts from the
ZAMS and it is followed until the end of core hydrogen burning and the
further  stages. Two  tracks are  shown, which  correspond  to initial
orbital periods  of 0.90 and 1.40d.  At the points denoted  by RL, the
secondary fills it  Roche Lobe for the first time,  and mass loss from
the system begins. Note that  from that moment on, the secondary evolves at
almost constant radius, until  its stellar mass decreases below $\sim$
0.4  $M_{\sun}$.  Mass loss  continues  until  the secondary  shrinks
within  its Roche  lobe.  The  final  mass at  the end  of Roche  lobe
overflow  is,  respectively,   0.15540  and  0.18685  $M_{\sun}$.  The
evolutionary  tracks for our He-core white  dwarf models  start precisely
after the end of Roche lobe overflow.

\begin{figure}
\centering
\includegraphics[clip,width=8.5cm ]{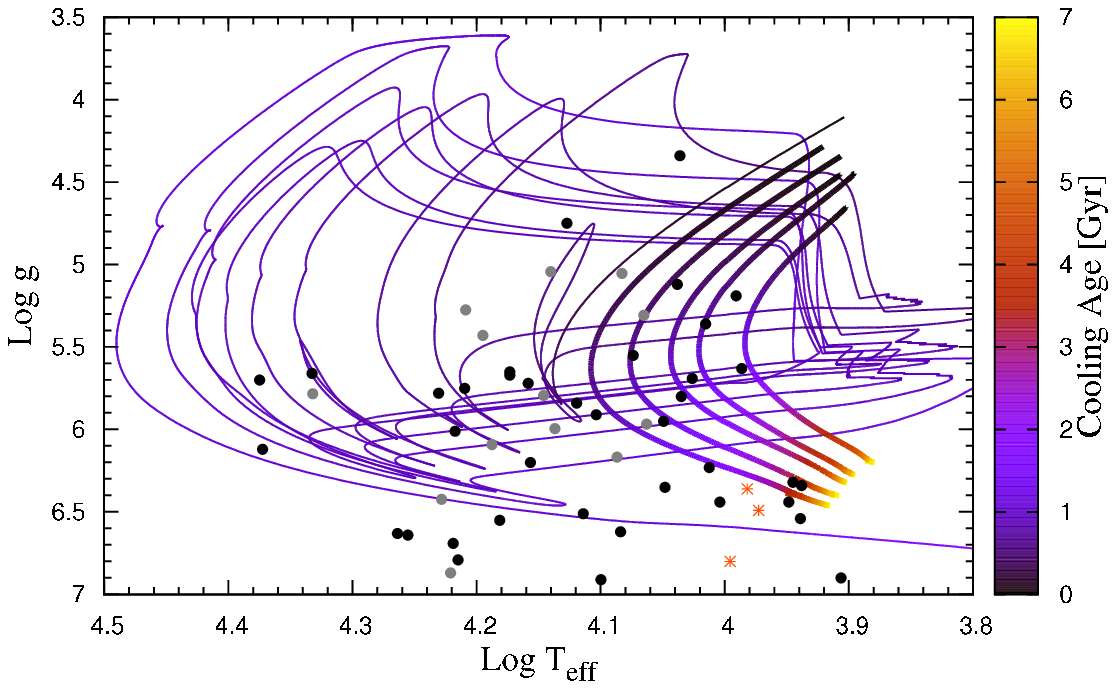}
\caption{Surface gravity - effective temperature diagram for ELM white
  dwarf sequences with 0.15540, 0.16115, 0.16499, 0.17064, 0.17624
  $M_{\sun}$ (thick lines, from right to left) together with the
  lowest mass He-core white dwarf sequence that undergoes CNO flashes 
  (0.18213 $M_{\sun}$, thin line). Tracks correspond to the first 7 Gyr 
   of evolution
  after the end of mass loss when the secondary shrinks within its
  Roche lobe.  The color scale to the right displays the cooling age in
  Gyr. Note the much faster evolution of the 0.18213 $M_{\sun}$ sequence 
  after the occurrence of the CNO flashes. Black and grey  circles 
  represent observed non-pulsating post-RGB low-mass stars, and red asteriks
  pulsating ELM white dwarfs, as detailed in Table \ref{Tab:ELMs}.}
\label{lowmass}
\end{figure}

The  evolution   in  the   plane  surface  gravity   versus  effective
temperature for all of our He-core  white dwarf sequences is  shown in
Fig. \ref{Fig:g-teff-ELMs}, together with selected post-RGB low-mass objects presumably
to be He-core white dwarfs (most of them ELM white dwarfs), see
Table \ref{Tab:ELMs}. 
In Fig. \ref{lowmass} we display on the same plot
the results  for the  ELM sequences which do  not experience
 CNO   flashes   on   the  cooling   branch, together with the
lowest mass He-core white dwarf sequence that experiences CNO flashes, 
the 0.18213 $M_{\sun}$ sequence (see   Table
\ref{tabla}).   The  figure   illustrates   the  evolutionary   stages
corresponding to  the first  7 Gyr  of evolution after the  end of mass
loss.  The main remarkable  characteristic of the no-flashing sequences 
is their slow evolution  over most of their  life, which is  due to the
residual  hydrogen shell  burning  being the  main  source of  surface
luminosity even at very  advanced stages of evolution.  In particular,
for the  least massive of  our ELM sequences,  evolutionary timescales
involved in  reaching the white dwarf  cooling branch from  the end of
mass loss amount to about 1 Gyr, which means that these objects should
have  large  chances  of  being  observed during  those  stages.  Also
noticeable  is  the  very  slow  rate  of  evolution  at  intermediate
effective  temperatures on  the  cooling branch.  In  fact, note  that
evolution  takes about  several  Gyr in  reaching  $T_{\rm eff}  \sim$
9000K. And  this is  so because even  at such advanced  stages, stable
hydrogen shell  burning still provides the  main energy source. It  is clear
that  for  these  ELM   white  dwarfs,  an  appropriate  treatment  of
progenitor  evolution is  required  for a  correct  assessment of  the
evolutionary time scales.

By contrast, the behavior is entirely different for the  sequences
that do experience unstable hydrogen shell burning on their early cooling
branch, as can appreciated by inspecting the track corresponding to
the 0.18213 $M_{\sun}$ sequence in Fig.  \ref{lowmass}.  This sequence
experiences nine CNO flashes before reaching the quiescent terminal
white-dwarf cooling branch. Several points are worthy of comment.  Note that
during the final cooling branch evolution proceeds on a timescale much
shorter than that characterizing the sequences with $M<0.18 M_{\sun}$.
This is because CNO flashes markedly reduce the hydrogen
content of the star, with the result that residual nuclear burning is
much less relevant when the remnant reaches the final cooling branch.
Hence,  sequences that go through
several episodes of CNO flashes are expected to evolve
below $T_{\rm eff} = 5000$K in a time much less than the Hubble time,
in contrast with the situation for the least massive ELM sequences for
which nuclear burning substantially delays their evolution by several
Gyr at higher effective temperatures.  This age dichotomy, which is
the result of the interplay of element diffusion and nuclear burning
during the flash episodes, is a remarkable property of very low mass
He-core white dwarfs with important observational consequences, as we
reported in previous investigations (see Althaus et al. 2001; Panei et al. 2007 
and references therein). 

\begin{figure}
\centering
\includegraphics[clip,width=8.5cm]{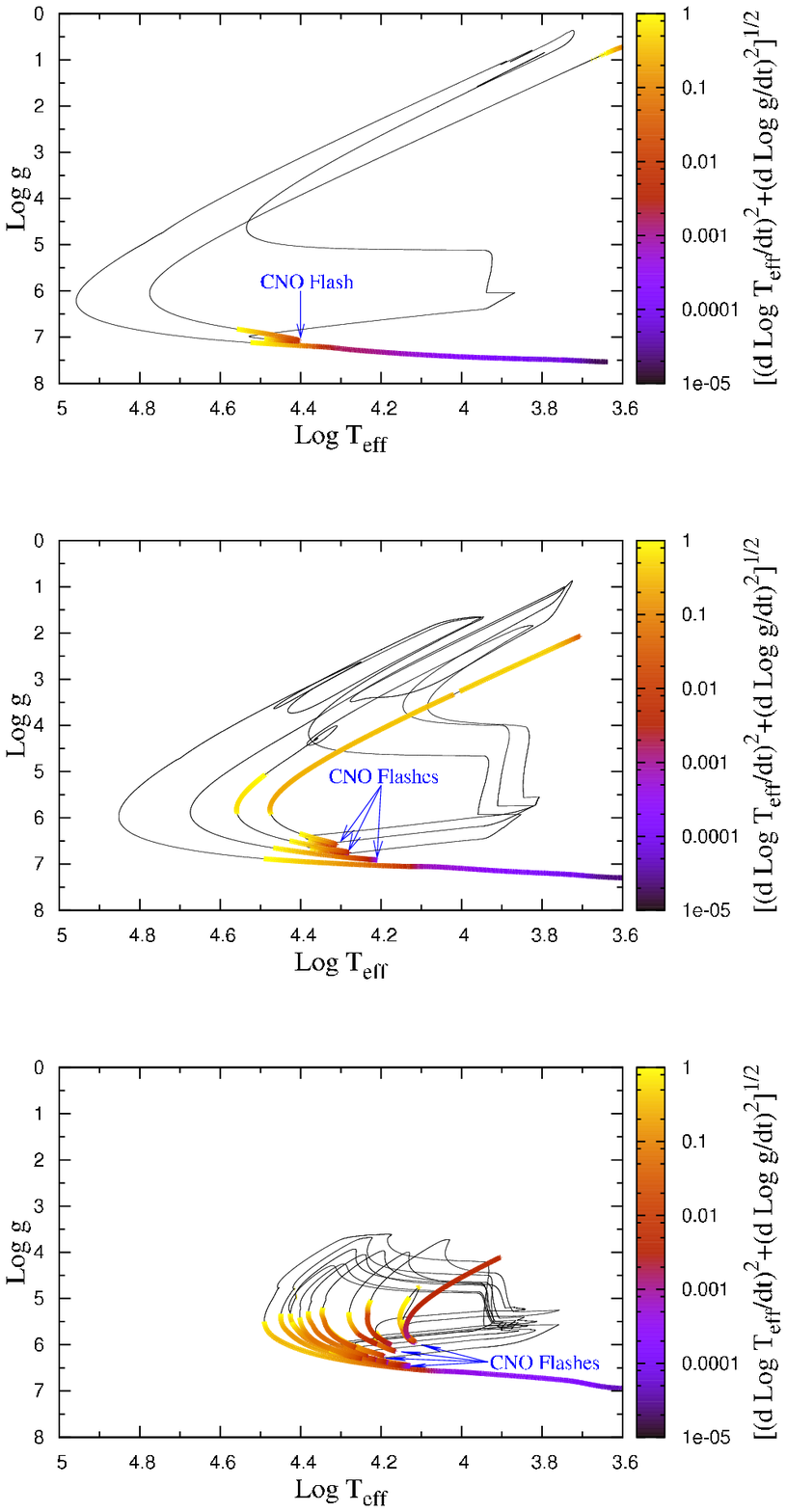}
\caption{Evolutionary speed of the sequences in the $\log T_{\rm
    eff}-\log g$ diagram ---for 0.36304 $M_{\sun}$, 0.27242, and
  0.18213$M_{\sun}$ (top, middle, and bottom panels,
  respectively). Thin black lines denote evolutionary stages faster
  than the upper boundary of the color coding.  Note the much faster
  evolution of the more massive sequences before entering the white
  dwarf cooling sequence.}
\label{Fig:Velocid}
\end{figure}

It is worth noting that, in the case of CNO-flashing sequences, the rate of
evolution slows down during the stages just prior to the occurrence of
the CNO flashes on the early cooling branches. This is illustrated
by Fig. \ref{Fig:Velocid}, which shows the evolutionary speed of some 
selected CNO-flashing sequences in the $\log T_{\rm eff}-\log g$ diagram. 
The slower evolution just before the 
occurrence of the CNO flashes is apparent. Hence, the
observational counterparts of these remnants could have chances of being
detected also during these evolutionary stages.  This should be taken
into account when attempt is made at assessing the stellar mass and
age of He-core white dwarfs from evolutionary sequences that
experience several CNO flashes, since multiple solutions are, in
principle, possible from such sequences, at intermediate effective
temperatures  (see Silvotti et al. 2012). This can be clearly
understood by inspecting Fig. \ref{Fig:g-teff-ELMs}.  We develop this
issue in the next section.

\section{Mass and Age determination}
\label{MassAge}

Our  grid  of  He-core   white  dwarf  sequences  is  appropriate  for
homegeneous  mass and cooling  age determinations  of observed  low-mass white
dwarfs for all of the  evolutionary stages where these stars have 
chances of being observed, i.e.,  the final cooling branch, the stages
at  constant  luminosity following  the  end  of  Roche lobe  overflow
(particularly in  the case of ELM  white dwarfs), and,  as we mention,
the  evolutionary  stages prior  to  the  occurrence  of the  CNO
flashes on  the early cooling  branches. These evolutionary stages are
highlighted in Fig.  \ref{Fig:Velocid} for some selected CNO-flashing
sequences.  Because  of the occurrence  of several
CNO flashes, special care must be  taken at assessing the age and mass
from theoretical He-core evolutionary sequences. Usually, white dwarf masses and
cooling ages  are derived  from their  $\log g$-$\log  T_{\rm eff}$  values by
means of  theoretical white dwarf  tracks and isochrones in  the $\log
g$-$\log  T_{\rm eff}$  plane. Performing  linear  interpolation among
available tracks and isochrones is usually enough to obtain masses and
cooling ages directly from the $\log g$-$\log T_{\rm eff}$ values. However, in
the  case of  low-mass  He-core  white dwarfs  the  problem is  rather
involved.   As can  be seen  in Fig.   \ref{Fig:g-teff-ELMs}  for high
effective temperatures, He-core white dwarf tracks cross each other in
the $\log  g$-$\log T_{\rm eff}$  plane, leading to a  multiplicity of
the possible  solutions, both for mass  and cooling age, for  a given measured
value for $\log  g$ and $\log T_{\rm eff}$.   This multiplicity of the
age and mass solutions would  be almost irrelevant if the evolutionary
stages  at which the  tracks cross  each other  (loops) were  fast and
consequently with  a low probability to  find a star  at those stages.
Unfortunately,   the   time   spent   during  these   loops   is   not
negligible\footnote{As we mentioned, most of  this time corresponds to  
the stages just prior  to the  occurrence of  the CNO  flashes, see 
Fig.  \ref{Fig:Velocid}.} and,  thus, several
solutions are  possible for  a given measured  value for $\log  g$ and
$\log  T_{\rm eff}$.   Consequently, the  determination of  masses and
ages of He-core  white dwarfs calls for some  kind of statistical approach
to weight  the time spent  by each model  in each region of  the $\log
g$-$\log T_{\rm eff}$ diagram.
 
\begin{figure}
\centering
\includegraphics[clip,width=8.5cm]{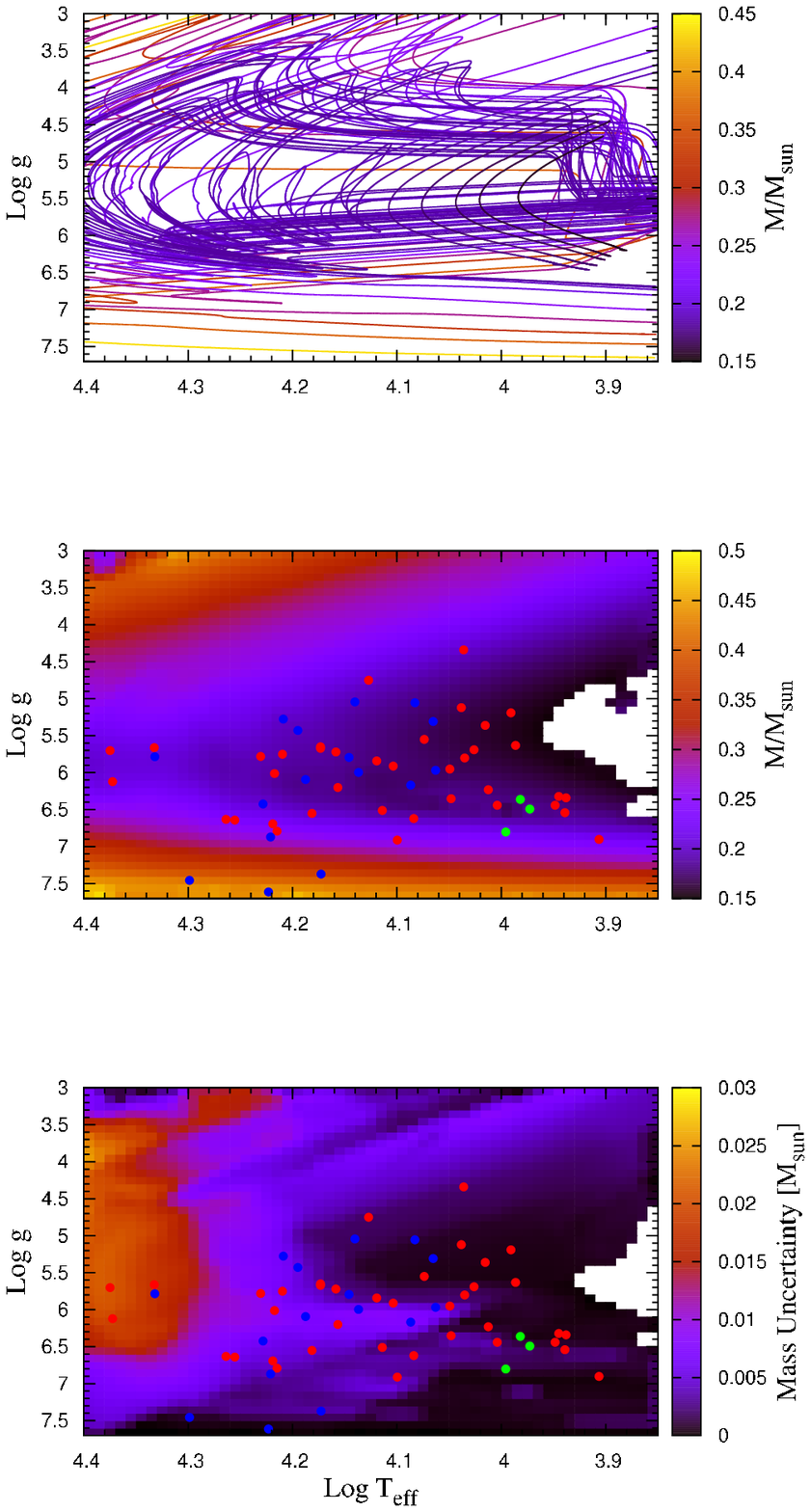}
\caption{Top Panel: $M(\log T_{\rm eff}, \log g)$ relation as obtained
  directly from stellar evolution simulations. Middle Panel: $M(\log
  T_{\rm eff}, \log g)$ relation derived from the weighted least
  square scheme described in Sect. \ref{MassAge}. Bottom Panel:
  Uncertainty $\sigma_M$ of the derived $M(\log T_{\rm eff}, \log g)$
  relation. Symbols correspond to the observed post-RGB low-mass stars listed
  in Table \ref{Tab:ELMs}. In particular, green circles correspond to
  the pulsating ELM white dwarfs detected by Hermes et
  al. (2013). White regions in middle and bottom panels indicate
   either fast evolutionary stages for which not enough points were
    available to obtain an accurate fit (in the case of CNO-flashing sequences), 
or simply that no tracks are available at that regions.}
\label{Fig:Masses}
\end{figure}
\begin{figure}
\centering
\includegraphics[clip,width=8.5cm]{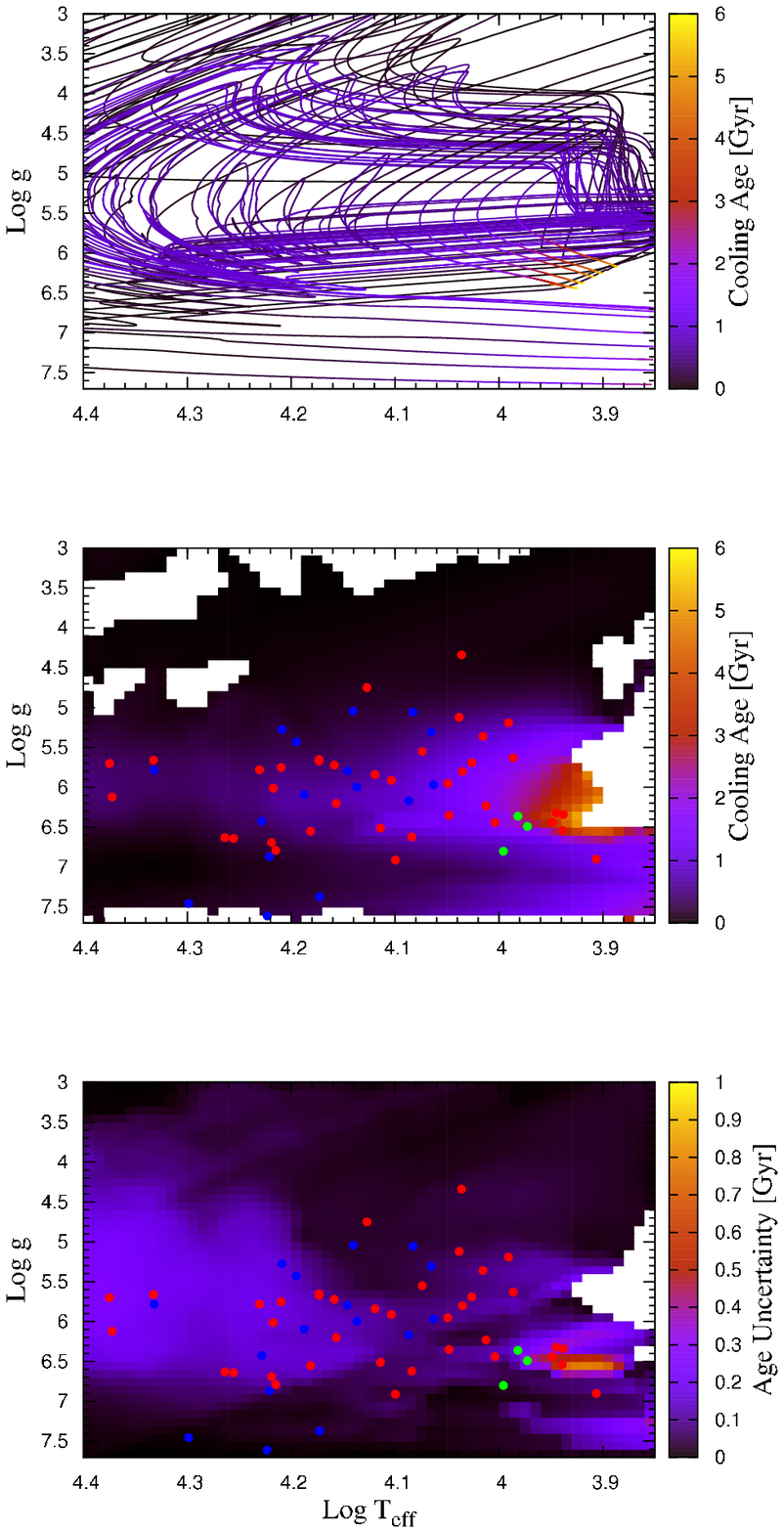}
\caption{Same as Fig. \ref{Fig:Masses} but for the $t(\log T_{\rm
    eff}, \log g)$ relation and its corresponding uncertainty
  $\sigma_t$.}
\label{Fig:Ages}
\end{figure}

In this connection, we adopted the following scheme to determine
masses and cooling ages of He-core white dwarfs from our sequences. In order to 
weight the time
spent by each model at different locations of the $\log g$-$\log
T_{\rm eff}$ diagram we created random values of cooling age $(t)$ and mass
($M_*$) and computed their corresponding theoretical values of $\log g$-$\log T_{\rm
  eff}$ from our sequences. Thus we are left with a set of points whose
 density in the $\log g$-$\log T_{\rm eff}$ diagram is directly
 related to the probability of finding a model of a given mass at that location of the 
 $\log g$-$\log T_{\rm eff}$ diagram. Because interpolation between
different tracks is not possible we adopted for each mass value that
of our closest model. In the absence of a better knowledge of He-core white dwarf progenitor 
lifetimes and masses, we assumed a uniform
distribution in mass and cooling age within the ranges $ M_*\in (0.14965
M_{\sun}, 0.50742 M_{\sun})$ and $t\in(0 {\rm Gyr},7 {\rm
  Gyr})$. Within these assumptions we computed $2\times 10^7$ points
to have a good sampling of the $\log g$-$\log T_{\rm eff}$-Mass-Age
relationship. Then, for each observed He-core white dwarf star with known values of
$\log g$-$\log T_{\rm eff}$ (see Table \ref{Tab:ELMs}) we performed a
least square polynomial fitting of the $\log g$-$\log T_{\rm
  eff}$-Mass-Age relationship within an ellipse around the ($\log
g^{\rm star}$,$\log T_{\rm eff}^{\rm star}$) values of the star and
derived its mass and age. We adopted the dispersion around the fitted
polynom as an estimation of the errors in mass and cooling age. The size of
the ellipse and the degree of the polynomial expression were chosen
to allow for a good representation of different regions of
the $\log g$-$\log T_{\rm eff}$ plane. Specifically, we adopted the
following expressions.  For the region where the ``knees'' of the tracks
are located ($\log g<-12.1+4.5\times \log T_{\rm eff}$) we adopted
\begin{eqnarray}
M(\log T_{\rm eff}, \log g)&=& a\, \log^2 T_{\rm eff}+b\,\log^2 g+c\, \log T_{\rm eff}\, \log g \nonumber\\
      &+&d\, \log T_{\rm eff}+e\, \log g+d.
\label{eq:M_tgcuad}
\end{eqnarray}
For values
of $\log g>-12.1+4.5\times \log T_{\rm eff}$, a first degree polynom
was adopted.

\begin{equation}
M(\log T_{\rm eff}, \log g)=a'\, \log T_{\rm eff}+b'\, \log g+c' 
\label{eq:M_Tglineal}
\end{equation}
Similar expressions were adopted in the same regions for the
$t(\log T_{\rm eff}, \log g)$ relation.  

To select the points to be used in each fitting we adopted
\begin{equation}
\left(\frac{\log T_{\rm eff}-\log T_{\rm eff}^{\rm
    star}}{0.04\times\alpha}\right)^2 +\left(\frac{\log g-\log g^{\rm
    star}}{0.6\times\beta}\right)^2\le 1
\end{equation}
Where $\alpha$ and $\beta$ have to be chosen in such a way to allow enough
points for the least square fit but, at the same time, keeping them small enough so
that the $t(\log T_{\rm eff}, \log g)$ and $M(\log T_{\rm eff}, \log
g)$ can be reproduced by the simple polynomial expressions of
Eq. \ref{eq:M_tgcuad} and \ref{eq:M_Tglineal}. In particular
 $\alpha$ and $\beta$ were taken as
\begin{eqnarray}
\alpha&=&1.5,\, \beta=1.5,\ \hbox{if}\ \log g<-34.6+9.4\times \log T_{\rm eff}\nonumber\\ 
\alpha&=&1,\, \beta=1.5, \nonumber\\
    &\hbox{if}& -34.6+9.4\times \log T_{\rm eff}\le \log g<-12.1+4.5\times \log T_{\rm eff}\nonumber\\ 
\alpha&=&1,\, \beta=0.5, \nonumber\\
&\hbox{if}& -12.1+4.5\times \log T_{\rm eff}\le \log g<-21.55+7\times \log T_{\rm eff}\nonumber\\ 
\alpha&=&1,\, \beta=0.325, \hbox{in any other case}.  
\end{eqnarray}

In Figs. \ref{Fig:Masses} and \ref{Fig:Ages} we can see the $M(\log
T_{\rm eff}, \log g)$ and $t(\log T_{\rm eff}, \log g)$ relations as given
directly by our stellar evolution sequences (top panels) together with the weighted 
least square fitted $M(\log T_{\rm eff}, \log g)$ and $t(\log T_{\rm eff},
\log g)$ relations described above (middle panels) and the estimated
uncertainties of the least squared fit procedure (bottom panel). It is worth
noting that the relatively large uncertainties of the interpolated
$M(\log T_{\rm eff}, \log g)$ relation (Fig. \ref{Fig:Masses}, bottom
panel) at $\log g\lesssim 6.5$ and $\log T_{\rm eff}\gtrsim 4.3$ is a
natural consequence of the fact that sequences of different masses
spend similar times in that region, making that region populated by
stars of different masses. Hence, multiple solutions (with a
dispersion of 0.03$M_{\sun}$) for the inferred stellar mass are
possible at that region. A similar situation can be appreciated in the
$t(\log T_{\rm eff}, \log g)$ relation at $5.5 \lesssim \log g\lesssim 6.5$ and
$\log T_{\rm eff}\gtrsim 4.1$ where the uncertainty of the derived
cooling ages is of about a few hundred Myrs. In addition, the $t(\log T_{\rm
  eff}, \log g)$ relation becomes very uncertain ($\sigma_t\sim
1$Gyr) around $\log g\sim 6.5$ and $3.9 \lesssim \log T_{\rm
  eff}\lesssim 3.95$. This feature is because at that region of the
$\log g$-$\log T_{\rm eff}$ diagram it is expected the age dichotomy
between the very low mass sequences ($M_*$$\lesssim$$0.18M_{\sun}$) which
do not undergo CNO flashes (and, as we saw, evolve at a very slow pace) 
and the higher mass sequences ($M_*$$\gtrsim$$0.18M_{\sun}$) which undergo CNO
flashes with the consequent much faster evolution during
the final cooling branch.

\begin{table*} 
\caption[]{Selected post-RGB low-mass stars. Masses and cooling ages (fourth and fifth columns) 
are calculated from
our sequences under the
assumption that they are He-core white dwarfs.  $\log g$ and
  $\log T_{\rm eff}$ values were taken from the tables of Silvotti et
  al. (2012, first 38 listed objects), Hermes et al. (2013, next 3
  listed objects) and Brown et al. (2013, final 17 listed objects). The last
two columns list the values as derived from our procedure
for quick mass and age determination, see Sect. \ref{quick}.}
\begin{tabular}{lcccc|cc}
\hline
Name              & T$_{\rm eff}$/K  & log $g$ (cm~s$^{-2})$ & M/M$_{{\sun}}$ & Cooling Age [Myr] &    M$^{\rm app}$/M$_{{\sun}}$ & C. Age$^{\rm app}$ [Myr] \\ 
\hline
V209~$\omega$~Cen & 10866 $\pm$ 323 & 4.34 $\pm$ 0.02 &{\bf 0.202 $\pm$ 0.0021 }& {\bf 55  $\pm$  31  }  &  0.206  & 85  \\ 
WASP~J0247$-$25   & 13400 $\pm$1200 & 4.75 $\pm$ 0.05 &{\bf 0.203 $\pm$ 0.0019 } &{\bf  110  $\pm$ 25  }  &  0.208  & 100  \\ 
SDSS~J1233+1602   & 10920 $\pm$ 160 & 5.12 $\pm$ 0.07 &{\bf 0.169 $\pm$ 0.0005 } &{\bf  434  $\pm$  50 }  &  0.172  & 547  \\ 
SDSS~J1741+6526   & 9790 $\pm$ 240  & 5.19 $\pm$ 0.06 &{\bf 0.159 $\pm$ 0.0003  }&{\bf  712  $\pm$  80 }  &  0.161 &  838   \\
SDSS~J2119$-$0018 & 10360 $\pm$ 230 & 5.36 $\pm$ 0.07 &{\bf 0.161 $\pm$ 0.0004 } &{\bf  885  $\pm$ 90  }  &  0.162 &  935   \\
SDSS~J0917+4638   & 11850 $\pm$ 170 & 5.55 $\pm$ 0.05 &{\bf 0.170 $\pm$ 0.0003 } &{\bf  703  $\pm$ 46   } &  0.172 & 762    \\     
SDSS~J0112+1835   & 9690 $\pm$ 150  & 5.63 $\pm$ 0.06 &{\bf 0.156 $\pm$ 0.0003 } &{\bf  1604 $\pm$  93  } &  0.158 &  1653  \\
KIC~10657664      & 14900 $\pm$ 300 & 5.65 $\pm$ 0.04 &{\bf 0.189 $\pm$ 0.0035 } &{\bf  382  $\pm$ 116  } &  0.189 & 421   \\ 
HD~188112         & 21500 $\pm$ 500 & 5.66 $\pm$ 0.05 &{\bf 0.211 $\pm$ 0.0175 } &{\bf  329  $\pm$ 207  } &  0.212 & 291  \\
GALLEX~J1717      & 14900 $\pm$ 200 & 5.67 $\pm$ 0.05 &{\bf 0.188 $\pm$ 0.0039  }&{\bf  406  $\pm$ 127  } &  0.188 & 432  \\ 
SDSS~J0818+3536   & 10620 $\pm$ 380 & 5.69 $\pm$ 0.07 &{\bf 0.163 $\pm$ 0.0014  }& {\bf 1263 $\pm$ 79   } &  0.165 & 1383  \\
KIC~06614501      & 23700 $\pm$ 500 & 5.70 $\pm$ 0.10 &{\bf 0.210 $\pm$ 0.0209 } &{\bf  368  $\pm$ 239  } &  0.217 & 287   \\
KOI~1224          & 14400 $\pm$1100 & 5.72 $\pm$ 0.05 &{\bf 0.186 $\pm$ 0.0024  }&{\bf  441  $\pm$ 82   } &  0.186 & 474    \\
NGC~6121~V46      & 16200 $\pm$ 550 & 5.75 $\pm$ 0.11 &{\bf 0.191 $\pm$ 0.0064 } & {\bf 450  $\pm$ 176  } &  0.192 &  435  \\
KOI~81            & 17000 $\pm$1300 & 5.78 $\pm$ 0.13 &{\bf 0.196 $\pm$ 0.0072 } &{\bf  372  $\pm$ 186  } &  0.194  & 427   \\
SDSS~J0152+0749   & 10840 $\pm$ 270 & 5.80 $\pm$ 0.06 &{\bf 0.167 $\pm$ 0.0020 } &{\bf  1312 $\pm$ 88   } &  0.167 &  1461  \\
SDSS~J0755+4906   & 13160 $\pm$ 260 & 5.84 $\pm$ 0.07 &{\bf 0.178 $\pm$ 0.0027 } &{\bf  633 $\pm$ 108   } &  0.179 & 687  \\
SDSS~J1422+4352   & 12690 $\pm$ 130 & 5.91 $\pm$ 0.07 &{\bf 0.177 $\pm$ 0.0025 } &{\bf  776 $\pm$ 95    } &  0.177 & 817   \\
SDSS~J1630+2712   & 11200 $\pm$ 350 & 5.95 $\pm$ 0.07 &{\bf 0.171 $\pm$ 0.0023 } &{\bf  1270 $\pm$ 89   } &  0.171 & 1302  \\
SDSS~J0106--1000  & 16490 $\pm$ 460 & 6.01 $\pm$ 0.04 &{\bf 0.186 $\pm$ 0.0099  }&{\bf  568 $\pm$ 174  }  &  0.193 & 524   \\
SDSS~J1625+3632   & 23570 $\pm$ 440 & 6.12 $\pm$ 0.03 &{\bf 0.214 $\pm$ 0.0188 } & {\bf 415 $\pm$ 212  }  &  0.220 & 304  \\
SDSS~J1439+1002   & 14340 $\pm$ 240 & 6.20 $\pm$ 0.07 &{\bf 0.182 $\pm$ 0.0070 } & {\bf 746 $\pm$ 125  }  &  0.186 & 705  \\
SDSS~J0849+0445   & 10290 $\pm$ 250 & 6.23 $\pm$ 0.08 &{\bf 0.178 $\pm$ 0.0003 } &{\bf  1780 $\pm$ 34   } &  0.176 & 1877  \\
SDSS~J1443+1509   &  8810 $\pm$ 220 & 6.32 $\pm$ 0.07 &{\bf 0.173 $\pm$ 0.0004 } &{\bf  3646 $\pm$ 200  } &  0.172 & 4037   \\
PSR~J1012+5307    & 8670 $\pm$ 300  & 6.34 $\pm$ 0.20 &{\bf 0.172 $\pm$ 0.0004  }&{\bf  4034 $\pm$ 169  } &  0.172 & 4341   \\
LP~400-22         & 11170 $\pm$ 90  & 6.35 $\pm$ 0.05 &{\bf 0.182 $\pm$ 0.0025 } &{\bf  1195 $\pm$ 100  } &  0.182 &  1279  \\
PSR~J1911$-$5958  & 10090 $\pm$ 150 & 6.44 $\pm$ 0.20 &{\bf 0.185 $\pm$ 0.0041 } &{\bf  1471 $\pm$ 167  } &  0.184 &  2324   \\
SDSS~J0822+2753   &  8880 $\pm$ 60  & 6.44 $\pm$ 0.11 &{\bf 0.178 $\pm$ 0.0010 } &{\bf  3255 $\pm$ 416  } &  0.179 &   3862   \\
KOI~74            & 13000 $\pm$1000 & 6.51 $\pm$ 0.10 &{\bf 0.186 $\pm$ 0.0016 } &{\bf  930  $\pm$ 64   } &  0.197 &  670   \\
NLTT~11748        & 8690 $\pm$ 140  & 6.54 $\pm$ 0.05 &{\bf 0.183 $\pm$ 0.0020 } &{\bf  2572 $\pm$ 507  } &  0.184 & 3011  \\
SDSS~J1053+5200   & 15180 $\pm$ 600 & 6.55 $\pm$ 0.09 &{\bf 0.204 $\pm$ 0.0045 } &{\bf  479 $\pm$ 147   } &  0.213 & 412  \\
SDSS~J1512+2615   & 12130 $\pm$ 210 & 6.62 $\pm$ 0.07 &{\bf 0.198 $\pm$ 0.0034 } &{\bf  750 $\pm$ 76   }  &  0.205 & 683  \\
SDSS~J0923+3028   & 18350 $\pm$ 290 & 6.63 $\pm$ 0.05 &{\bf 0.238 $\pm$ 0.0067 } &{\bf  188 $\pm$ 112  }  &  0.246 & 178  \\
SDSS~J1234$-$0228 & 18000 $\pm$ 170 & 6.64 $\pm$ 0.03 &{\bf 0.235 $\pm$ 0.0063 } &{\bf  225 $\pm$ 124  }  &  0.245 & 194   \\
SDSS~J1436+5010   & 16550 $\pm$ 260 & 6.69 $\pm$ 0.07 &{\bf 0.234 $\pm$ 0.0047 } &{\bf  230 $\pm$ 126  }  &  0.240 &  264  \\
SDSS~J0651+2844   & 16400 $\pm$ 300 & 6.79 $\pm$ 0.04 &{\bf 0.248 $\pm$ 0.0041 } &{\bf  196 $\pm$ 107  }  &  0.260 & 262   \\
PSR~J0218+4232    & 8060 $\pm$ 150  & 6.90 $\pm$ 0.70 &{\bf 0.227 $\pm$ 0.0001 } &{\bf  1162 $\pm$ 48  }  &  0.237 & 1505   \\
SDSS~J1448+1342   & 12580 $\pm$ 230 & 6.91 $\pm$ 0.07 &{\bf 0.250 $\pm$ 0.0030 } &{\bf  302 $\pm$ 13   }  &  0.263 & 525   \\\hline
SDSS~J1112+1117   & 9590 $\pm$ 140  & 6.36 $\pm$ 0.06 &{\bf 0.179 $\pm$ 0.0012 } &{\bf  2756 $\pm$ 218 }  &  0.178 &  3061  \\
SDSS~J1840+6423   & 9390 $\pm$ 140  & 6.49 $\pm$ 0.06 &{\bf 0.183 $\pm$ 0.0018 } &{\bf  1926 $\pm$ 453  } &  0.184 &  2766  \\
SDSS~J1518+0658   & 9900 $\pm$ 140  & 6.80 $\pm$ 0.05 &{\bf 0.220 $\pm$ 0.0013 } &{\bf  733 $\pm$ 33    } &  0.224 &  829   \\\hline
J0840+1527 & 13810 $\pm$  240 &  5.043  $\pm$  0.053 &{\bf 0.193 $\pm$ 0.0018 } &{\bf  181 $\pm$  23  } & 0.200 &  199   \\
J1157+0546 & 12100 $\pm$  250 &  5.054  $\pm$  0.071 &{\bf 0.180 $\pm$ 0.0008 } &{\bf  263 $\pm$  32  } & 0.185 &  323   \\
J1238+1946 & 16170 $\pm$  260 &  5.275  $\pm$  0.051 &{\bf 0.199 $\pm$ 0.0064 } &{\bf  233 $\pm$ 157  } & 0.205 &  225   \\
J1141+3850 & 11620 $\pm$  200 &  5.307  $\pm$  0.054 &{\bf 0.171 $\pm$ 0.0005 } &{\bf  526 $\pm$  55  } & 0.173 &  584   \\
J0751-0141 & 15660 $\pm$  240 &  5.429  $\pm$  0.046 &{\bf 0.200 $\pm$ 0.0053 } &{\bf  223 $\pm$ 137 }  & 0.196 &  287   \\
J0815+2309 & 21470 $\pm$  340 &  5.783  $\pm$  0.046 &{\bf 0.207 $\pm$ 0.0171 } &{\bf  373 $\pm$ 198  } & 0.208 &  333   \\
J0811+0225 & 13990 $\pm$  230 &  5.794  $\pm$  0.054 &{\bf 0.183 $\pm$ 0.0028 } &{\bf  528 $\pm$ 104 }  & 0.183 &  542   \\
J1538+0252 & 11560 $\pm$  220 &  5.967  $\pm$  0.053 &{\bf 0.172 $\pm$ 0.0056 } &{\bf  1155 $\pm$ 89 }  & 0.173 &  1184  \\
J2132+0754 & 13700 $\pm$  210 &  5.995  $\pm$  0.045 &{\bf 0.176 $\pm$ 0.0080 } &{\bf  732 $\pm$ 125  } & 0.181 &   725  \\
J1151+5858 & 15400 $\pm$  300 &  6.092  $\pm$  0.057 &{\bf 0.183 $\pm$ 0.0072 } &{\bf  647 $\pm$ 147  } & 0.188 &  628   \\
J0056-0611 & 12210 $\pm$  180 &  6.167  $\pm$  0.044 &{\bf 0.175 $\pm$ 0.0078 } &{\bf  982 $\pm$ 77  } &  0.176 &  1055  \\
J0802-0955 & 16910 $\pm$  280 &  6.423  $\pm$  0.048 &{\bf 0.206 $\pm$ 0.0063 } &{\bf  361 $\pm$ 158 }  & 0.213 &  347   \\
J2338-2052 & 16630 $\pm$  280 &  6.869  $\pm$  0.050 &{\bf 0.263 $\pm$ 0.0040  }&{\bf  168 $\pm$  99  } & 0.279 &   242  \\
J1046-0153 & 14880 $\pm$  230 &  7.370  $\pm$  0.045 &{\bf 0.376 $\pm$ 0.0055 } &{\bf  224 $\pm$  29 }  & 0.381 &  269   \\
J0755+4800 & 19890 $\pm$  350 &  7.455  $\pm$  0.057 &{\bf 0.422 $\pm$ 0.0008 } &{\bf  104 $\pm$  3  } &  0.420 &   99    \\
J1104+0918 & 16710 $\pm$  250 &  7.611  $\pm$  0.049 &{\bf 0.455 $\pm$ 0.0001 } &{\bf  158 $\pm$   3  } & 0.442 &  182    \\
J1557+2823 & 12550 $\pm$  200 &  7.762  $\pm$  0.046 &{\bf 0.456 $\pm$ $<$0.0001 } &{\bf  541 $\pm$ 1 } & 0.460 &  415    \\
\end{tabular}

\label{Tab:ELMs}
\end{table*}

As an application of our He-core white dwarf sequences and the above
described interpolation algorithm, we have determined the stellar
masses and cooling ages for the sample of post-RGB low-mass  stars (most
of them ELM white dwarfs) listed in Silvotti et al. (2012) and in Brown et al. (2013).  
We assume that these stars have a He core. The derived values of mass
and cooling age for each star is shown in Table \ref{Tab:ELMs} together with
the estimated uncertainties (fourth and fifth columns). We mention that
for such white dwarfs, the age values listed in Table \ref{Tab:ELMs}
constitute the first assessment of their cooling ages. As a check, we
compare our derived values  with those derived by the
standard interpolation approach, which can be performed for white dwarfs
with masses $M_*<0.18M_{\sun}$\footnote{Theoretical sequences with
  masses $M_*<0.18M_{\sun}$ do not cross each other and the sequences
  with masses $M_*>0.18M_{\sun}$ spend a negligible time in the region
  of the tracks with lower masses. Thus no degeneracy of the possible
  solutions exist, making possible to use a standard approach to
  determine masses and cooling ages of ELM-white dwarfs.}, and found that the difference
in the masses and ages is well within the quoted uncertainties.  We
want to mention that some differences appear between the inferred
stellar mass from our sequences and stellar mass values quoted by
Silvotti et al (2012) and Brown et al. (2013).  This is particularly true in the case
of the ELM white dwarf sample of Brown et al. (2013), who, for most of
their sample, found a stellar mass value of $0.17M_{\sun}$, in
contrat with our analysis, which yields stellar mass values in the range 
0.17$M_{\sun}$ - 0.21$M_{\sun}$. This difference is due to the fact that Brown et
al. (2013) adopted the final cooling track (after all the flashes) as
the representative one for the sequences which undergo CNO-flashes. As
shown in Fig. \ref{Fig:Velocid}, it is the first entrance to the white
dwarf cooling phase which is slower and takes longer time. Thus, for
sequences which undergo CNO flashes it is more probable to find a star
of a given mass during its first entrance (which occurs at lower
effective temperatures) than during its final entrance to the white
dwarf cooling zone. Choosing the final cooling track to derive masses
at  gravities $\log g\lesssim 6$ leads
to an underestimation of the stellar mass of up to 20\%. Note that the
differences in the evolutionary speeds of different sequences is 
naturally taken into account in the scheme
presented in the previous section.

We also checked the estimations performed by the scheme presented here
by comparing our results with the analysis performed by Silvotti et
al. (2012) on KIC 6614501. By assuming that the binary component is a He-core
white dwarf they discussed two possibilities: i) its mass  is
$M_1\sim0.185 M_\odot$ and the star is on the final cooling track
after all the CNO flashes, which occurs at a faster pace and it is
thus less probable (see Fig. \ref{Fig:Velocid}.) and ii) its mass is $M_1\gtrsim0.19 M_\odot$ 
and the star is still on one
of the cooling branches during the CNO-flash stage. In the second case
the mass of the star should be constrained between 0.19$-$0.24
$M_\odot$, from a comparison with Althaus et al. (2001) tracks and the
cooling age should be between 30 and 290 Myr. On the first, and less
probable case, the age derived by Silvotti et al. (2012) is of 180
Myr. By looking at Table \ref{Tab:ELMs} we see that our mass and cooling age
determination scheme gives a value of $M_1=0.210\pm 0.0209
M_\odot$ and $t_{\rm age}=368\pm 239$ Myr for KIC 6614501. Then
our mass derivation is in very good agreement with the values derived
for option ii of Silvotti et al. (2012) and the same happens for the cooling age of the
star. Note, in particular, that the large derived uncertainty in the
mass and age within our approach is due to the fact that many
different tracks go through that region of the $\log g$-$\log T_{\rm
  eff}$ plane at similar paces and, consequently a better
determination of mass and age just from $\log g$ and $\log T_{\rm
  eff}$ is not possible.

\subsection{A quick way to assess  mass and cooling age of He-core white dwarfs}
\label{quick}

In order to provide other authors with an easy way to estimate the
masses and cooling ages of He-core white dwarf stars from our tracks, we computed $M_*$ 
and $t$ with the previously described algorithm at a small grid of points in the
$\log g$-$\log T_{\rm eff}$ diagram. Such a grid is presented in Table
\ref{Tab:fast} and can be used via direct bilinear interpolation
to obtain the masses and cooling ages at any point in the $\log g$-$\log
T_{\rm eff}$ diagram. Because the grid is not regularly spaced in
$\log T_{\rm eff}$ (for the sake of economy), interpolation must be
performed first in $\log T_{\rm eff}$ and then in $\log g$. As an example,
the results from  these approximated estimations are listed in columns sixth and seventh of Table
\ref{Tab:ELMs} and compared with those directly computed from the scheme presented
in the previous section. While some differences are present in the
derived cooling ages between the full scheme and the bilinear interpolation of
Table \ref{Tab:fast}, these differences are moderate and similar to
the quoted uncertainties. In particular, for the ELM KIC
6614501 analysed by Silvotti et al. (2012), a simple bilinear 
interpolation from Table \ref{Tab:fast}
gives straightforwardly $M_1\sim$0.217 $M_\odot$ and $t_{\rm age}\sim$287 Myr, very
similar to the values derived by Silvotti et al. (2012). 
On the other hand, masses derived by the
bilinear interpolation of the grid presented in Table \ref{Tab:fast}
are in excellent agreement by those derived from the full
scheme. Globally, bilinear interpolation of the values presented in
Table \ref{Tab:fast} allows a very fast and easy way to derive masses
and cooling ages for He-core white dwarfs. It should be noted however that, in the case of the
age derivation, the bilinear interpolation within the rather coarse
grid of Table \ref{Tab:fast} does not offer a good representation for
those stars close to the age dichotomy that appears at higher
gravities and lower temperatures. For those stars a more precise cooling age
and mass derivation as those presented in Sect. \ref{MassAge}  can be
inferred from direct interpolation of the terminal cooling branch
of our evolutionary sequences.

\section{Prospects for an asteroseismic tool}
\label{pulsation}

\begin{table} 
\caption[]{Masses and cooling ages at a grid in the log($g$)-log T$_{\rm eff}$ plane.}
\begin{tabular}{cc|cc}
\hline
log T$_{\rm eff}$/K  & log($g$/cm~s$^{-2})$ & M/M$_{{\sun}}$ & Cooling Age [Myr] \\\hline\hline
      3.92   &     4.00    &    0.191   $\pm$     0.0019  &     19  $\pm$     27 \\
      4.00  &      4.00   &     0.215   $\pm$     0.0032   &    54  $\pm$     24 \\
      4.03   &     4.00   &     0.225   $\pm$     0.0038   &    56  $\pm$     14 \\
      4.20   &     4.00   &     0.278   $\pm$     0.0041  &     8  $\pm$     25 \\
      4.40   &     4.00   &     0.346   $\pm$     0.0267   &     3  $\pm$    125 \\\hline
      3.92   &     4.70   &     0.158   $\pm$     0.0004   &   109  $\pm$     21 \\
      4.00   &     4.70   &     0.174   $\pm$     0.0009  &    122  $\pm$     33 \\
      4.03  &      4.70   &     0.181   $\pm$     0.0013  &    119  $\pm$     30 \\
      4.20  &      4.70   &     0.232   $\pm$     0.0056  &     51  $\pm$     76 \\
      4.50   &     4.70   &     0.296   $\pm$     0.0257  &     42  $\pm$    155 \\\hline
      3.99  &      5.40   &     0.156  $\pm$      0.0003 &    1151 $\pm$    115 \\
      4.10   &     5.40   &     0.176   $\pm$     0.0004  &    430  $\pm$     34 \\
      4.15   &     5.40   &     0.187   $\pm$     0.0016 &     302  $\pm$     49 \\
      4.30   &     5.40   &     0.218   $\pm$     0.0160 &     202  $\pm$    200 \\
      4.50   &     5.40   &     0.242   $\pm$     0.0307 &     202  $\pm$    241 \\\hline
      3.96  &      5.80   &     0.156   $\pm$     0.0007 &    2300  $\pm$     53 \\
      4.10  &      5.80   &     0.176   $\pm$     0.0020 &     734  $\pm$     76 \\
      4.15  &      5.80   &     0.184   $\pm$     0.0029 &     528  $\pm$    105 \\
      4.30  &      5.80   &     0.202   $\pm$     0.0152  &    356  $\pm$     187 \\
      4.50  &      5.80   &     0.235   $\pm$     0.0301  &     245  $\pm$    243 \\\hline
      3.90  &      6.10   &     0.155   $\pm$     0.0002  &    4532  $\pm$    129 \\
      4.05  &      6.10   &     0.173   $\pm$     0.0064  &    1284  $\pm$     86 \\
      4.15  &      6.10   &     0.182   $\pm$     0.0079  &    727  $\pm$    133 \\
      4.30  &      6.10   &     0.207   $\pm$     0.0129 &     345  $\pm$    156 \\
      4.50 &       6.10   &     0.240   $\pm$     0.0295 &     255  $\pm$    233 \\\hline
      3.91 &       6.25   &     0.163   $\pm$     0.0002 &    4848  $\pm$    124 \\
      4.00 &       6.25   &     0.177   $\pm$     0.0004  &   1980  $\pm$     48 \\
      4.10 &       6.25   &     0.175   $\pm$     0.0074  &    879 $\pm$     89 \\
      4.30 &       6.25   &     0.219   $\pm$     0.0119  &    256  $\pm$   159 \\
      4.50 &       6.25   &     0.249   $\pm$     0.0286  &    236  $\pm$    233 \\\hline
      3.93 &       6.40   &     0.174   $\pm$     0.0004  &   4847  $\pm$    129 \\
      4.05 &       6.40   &     0.185   $\pm$     0.0052  &   1132  $\pm$    125 \\
      4.15 &       6.40    &    0.187   $\pm$     0.0056  &    547  $\pm$    152 \\
      4.30 &       6.40   &     0.232   $\pm$     0.0108  &    181  $\pm$    149 \\
      4.50 &       6.40   &     0.262   $\pm$     0.0230   &   157  $\pm$    184 \\\hline
      3.80 &       6.70   &     0.180   $\pm$     0.0006 &    2222  $\pm$     22 \\ 
      4.00 &        6.70  &     0.202   $\pm$     0.0018 &     799  $\pm$     49 \\
      4.15 &       6.70   &     0.221   $\pm$     0.0032  &    429  $\pm$      62 \\
      4.30 &       6.70   &     0.264   $\pm$     0.0086  &     60  $\pm$    86 \\
      4.50 &       6.70    &    0.310   $\pm$     0.0076  &     51  $\pm$    32 \\\hline
      3.80 &       7.43    &    0.347   $\pm$     0.0020  &   2546  $\pm$    102 \\
      4.00 &       7.43   &     0.369   $\pm$     0.0033  &    786  $\pm$   107 \\ 
      4.15 &       7.43    &    0.391   $\pm$     0.0031  &    289  $\pm$    31 \\
      4.30 &       7.43   &     0.415   $\pm$     0.0009  &    96  $\pm$     3 \\
      4.40 &       7.43    &    0.435   $\pm$     0.0050  &    45  $\pm$     8 \\
\end{tabular}
\label{Tab:fast}
\end{table}

\begin{figure}
\centering
\includegraphics[clip,width=8.5cm]{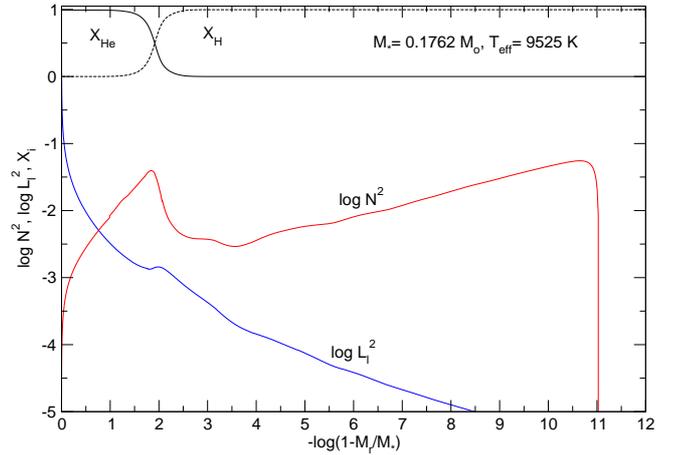}
\caption{The internal chemical profiles of He and H  and
the propagation diagram ---the run of the logarithm of the 
squared Brunt-V\"ais\"al\"a and Lamb (with $\ell= 1$) frequencies---  
corresponding to an $0.1762 M_{\sun}$ ELM white dwarf model at 
$T_{\rm eff}$$\sim$$9\,500$ K.}
\label{Fig:profiles-bvf-01762}
\end{figure}

\begin{figure}
\centering
\includegraphics[clip,width=8.5cm]{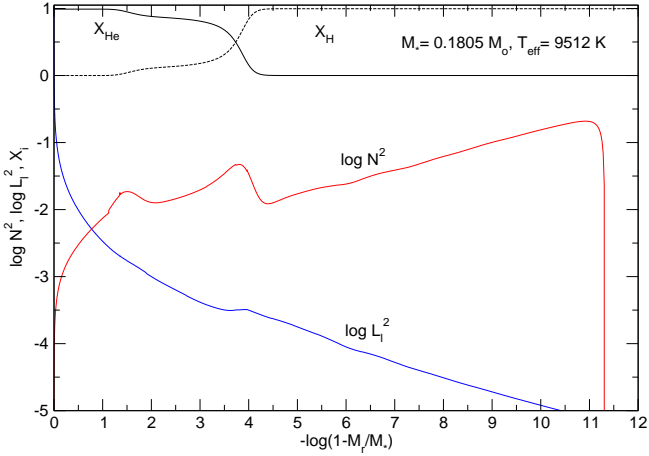}
\caption{Same as in Fig. \ref{Fig:profiles-bvf-01762}, 
but for a $0.1805 M_{\sun}$  ELM white dwarf model.}
\label{Fig:profiles-bvf-01805}
\end{figure}

We have carried out exploratory computations of the $g$-mode adiabatic
pulsation properties of our ELM  white dwarf models, in particular for
sequences  with  stellar  masses   near  the  critical  mass  for  the
development of  CNO flashes ($M_*\approx 0.2M_{\sun}$).  This is motivated
by the fact that at least  two of the three ELM pulsating white dwarfs
reported  by  Hermes et  al.  (2013)  have  stellar masses  near  this
threshold value. A full  exploration of the adiabatic and nonadiabatic
pulsation properties  of our  complete set of  low-mass and  ELM white
dwarfs  is underway  and  will be  presented  in a  future paper.  Our
pulsation  computations have  been performed  with  the Newton-Raphson
nonradial   pulsation   code  described   in   C\'orsico  \&   Althaus
(2006). Here, we focus on the most massive ELM sequence which does not
experience CNO flashes on  the cooling branch ($M_*=0.1762 M_{\sun}$),
and  the lowest  mass ELM  white dwarf  sequence that  experiences CNO
flashes        ($M_*=0.1805         M_{\sun}$).        In        Figs.
\ref{Fig:profiles-bvf-01762}   and   \ref{Fig:profiles-bvf-01805}   we
display  the chemical profiles  and the  propagation diagrams  for two
models  at $T_{\rm eff}\sim9\,500$  K.  It  is quite  noteworthy that,
being the  difference in the stellar  mass of these  two models almost
negligible  ($\Delta  M_*=0.0043  M_{\sun}$),  the  internal  chemical
distribution  is  very  different.   In  particular,  the  $M_*=0.1762
M_{\sun}$ model  has a {\it pure}  H envelope that is  $\sim 65$ times
thicker than the $M_*=0.1805 M_{\sun}$  model.  Also, the shape of the
He/H transition  region is markedly different in  both cases.  Indeed,
for  the   more  massive  model  this  chemical   interface  is  still
characterized  by  a  double-layered  structure, while  for  the  less
massive model  it has already attained a  single-layered structure due
to the  action of element diffusion. This  markedly different behavior
is  a  signature of  the  occurrence of  CNO  flashes  in the  $0.1805
M_{\sun}$   sequence,    and   can   be    understood   by   examining
Fig. \ref{lowmass}. In fact, as a result of the less important role of
residual    H   burning    after   the    occurrence   of    the   CNO
flashes\footnote{Because  of  the   various  CNO-flash  episodes,  the
  0.18213 $M_{\sun}$ sequence enters its final cooling branch with a H
  content a  factor 2-3 smaller,  thus making residual H  burning less
  relevant  on  the final  cooling  branch.},  the 0.18213  $M_{\sun}$
sequence   evolves   much    faster   than   the   $0.1762   M_{\sun}$
sequence.  Hence,  the  chemical  profile in  the  0.18213  $M_{\sun}$
sequence is not  expected to be completely separated  out by diffusion
processes, as in the case of the less massive sequence.

The differences in the shape of the He/H interface for both models are
translated into distinct  features in the run of  the squared critical
frequencies, in  particular in  the Brunt-V\"ais\"al\"a frequency ($N^2$),  
as can be appreciated  in Figs.
\ref{Fig:profiles-bvf-01762}  and  \ref{Fig:profiles-bvf-01805}, which
depict also the  propagation  diagrams  (Unno  et  al.   1989)
corresponding to these models. In fact, for the $M_*=0.1805 M_{\sun}$
model, $N^2$  is characterized by two  bumps, as it is  expected for a
double-layered chemical structure at the He/H.  At variance with this,
for the $M_*=0.1762 M_{\sun}$ model the Brunt-V\"ais\"al\"a frequency
has only  one bump, located  at $\log(1-M_r/M_*)\sim-1.9$,  which is
notoriously more  pronounced than its counterpart in  the $M_*=0.1805
M_{\sun}$ model.

\begin{figure}
\centering
\includegraphics[clip,width=8.5cm]{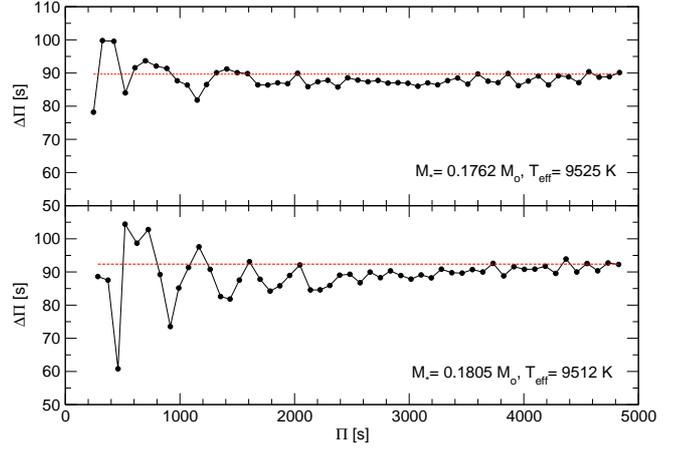}
\caption{The forward period spacing  in terms of the ($\ell$=$1$) periods 
for the model with $M_*$=$0.1762 M_{\sun}$ (upper panel) and 
$M_*$=$0.1805 M_{\sun}$ (lower panel), and a $T_{\rm eff}$$\sim$$9\, 500$ K. 
The red horizontal line corresponds to the asymptotic period spacing.}
\label{Fig:delp}
\end{figure}

In pulsating stars in general, the structure of the $g$-mode period 
spectrum is  sensitive to the spatial run of the Brunt-V\"ais\"al\"a 
frequency. This is particularly true for pulsating white dwarfs, in which
the bumps of $N^2$ induced by chemical gradients are responsible for 
mode-trapping effects (Bradley 1996; C\'orsico et al. 2002). 
A clear signature of mode trapping  is that the forward period
spacing, defined as $\Delta \Pi_k (\equiv \Pi_{k+1} - \Pi_k$), 
when plotted in terms of the pulsation period $\Pi_k$, 
exhibits strong departures from uniformity. The period difference 
between an observed mode and adjacent modes ($\Delta k \pm 1$) 
can be used as an observational diagnostic of mode trapping. 
For stellar models characterized by a single chemical interface, 
like the one of $M_*=0.1762 M_{\sun}$ we are considering here, 
local minima in $\Delta \Pi_k$ 
usually correspond to modes trapped in the H envelope, whereas local 
maxima in $\Delta \Pi_k$  are associated to modes trapped in the core 
region. In Fig. \ref{Fig:delp} we show the forward period spacing  in 
terms of the dipole periods for the same models with $M_*=0.1762 M_{\sun}$ 
and $M_*=0.1805 M_{\sun}$ described in Figs. \ref{Fig:profiles-bvf-01762}  
and  \ref{Fig:profiles-bvf-01805}. We depict with a red horizontal line
the asymptotic period spacing, computed as in C\'orsico et al. (2012).
From this figure, mode-trapping signatures are clearly 
noticeably for both models, particularly for periods shorter 
than $\sim 2500$ s. Longer periods seem to fit the asymptotic predictions,
although small departures from constant period spacing are still appreciable.

Focusing on the non-asymptotic regime ($\Pi_k \lesssim 2500$ s), we found 
appreciable differences in the period-spacing distributions of the 
two models. This, in spite of the fact that both models have virtually the 
same stellar mass. In particular, for the $M_*=0.1805 M_{\sun}$ model
the amplitude of the departure from  uniform period spacing due to 
mode trapping is appreciably larger than for the $M_*=0.1762 M_{\sun}$ 
one. Notably, this could be thought  as a useful seismic tool for 
discriminating stars that
have undergone CNO flashes in their early cooling phase from 
those that have not, provided that enough consecutive 
low- and intermediate-order ($200 \lesssim \Pi \lesssim 2000$ s) 
$g$-modes with the same $\ell$ value were detected in pulsating ELM white dwarfs. 
At least two of the pulsating ELM white dwarfs reported by Hermes et al. (2013), 
with stellar mass
near the threshold value for the occurrence of CNO flashes are potential candidates 
to test these ideas.

\section{Conclusions}
\label{conclusion}

In view of the discovery of numerous Extremely Low  Mass (ELM) white dwarfs
from  different  surveys  (Brown  et  al. 2012,  2013  and  references
therein)  and the  recent  detection  of pulsations  in  some of  them
(Hermes et al.  2012, 2013), we present in this  paper a detailed grid
of  evolutionary  sequences  for He-core  white dwarfs  by
considering the binary evolution  that leads to their formation. These
new  evolutionary models  are intended  for homogeneous  mass  and cooling age
determinations   as   well   as  for   potential   asteroseismological
applications of these stars.

Evolutionary calculations have been done  with the amply used and well
tested {\tt LPCODE} stellar evolutionary code, appropriately modified to
simulate the  binary evolution of progenitor stars.  Binary evolution 
has  been assumed to be fully  non-conservative, and the
loss  of  angular  momemtum  due  to  mass  loss,  gravitational  wave
radiation and magnetic  braking has been assumed.  All of our  He-core white 
dwarf initial models have been derived
from evolutionary calculations for binary systems consisting of an
evolving low-mass component of initially $1\, M_{\sun}$ and a $1.4\, M_{\sun}$ 
neutron star as the other component. Metallicity
is assumed to be Z=0.01. A total
of 14 initial  He-core white dwarf models with  stellar masses between
0.155  and 0.435  $M_{\sun}$  have been  computed  for initial  orbital
periods at the  beginning of the Roche lobe phase in  the range 0.9 to
300 d, see Table \ref{tabla}. In particular 9 sequences span the range
of masses corresponding to ELM-white dwarfs ($M\lesssim 0.20 M_{\sun}$).
It should be mentioned that in this paper we have not considered different values of
the initial mass of the secondary (mass-losing) star. Although it is  expected
that  the envelope mass of
the resulting white dwarf, a key factor in dictating the cooling times, is only
weakly dependent on the initial mass of the mass-losing star (Nelson
et al. 2004), different  orbital angular-momentum loss prescriptions due 
to mass loss  could have
an impact on the final envelope mass. This issue  has not been considered in this paper. 
Neither have
we explored other physical processes, such as pulsar irradiation, which could 
reduce the resulting envelope mass of the He-core white dwarf (Ergma et al. 2001).

In agreement with  previous studies, we found that  for stellar masses
larger than about 0.18 $M_{\sun}$, multiple CNO flashes are expected.
Because of this,  special care must be taken at  assessing the cooling age and
mass  from evolutionary  sequences, since  multiple solutions  are, in
principle,  possible from  such sequences,  at  intermediate effective
temperatures.  To  get realiable mass  and cooling age determination  from our
sequences (given  the surface  gravity and effective  temperature), we
devised  an interpolation  algorithm that  accounts for  precisely the
several possible solutions for stellar mass and cooling age in the case of the
CNO-flashing sequences.  As an application of our He-core white dwarf
sequences, we have determined the  stellar masses and cooling ages for
the  sample  of  post RGB low-mass stars  considered  in  Silvotti  et
al.  (2012) and  Brown et  al. (2013).  Finally, we  provide  a simple
scheme  with   the  aim  that   our  evolutionary  sequences   can  be
straightforwardly used  for mass and  age inferences of  He-core white
dwarfs.

Finally, we have explored the pulsation properties of ELM
white dwarf models for sequences with stellar masses near the critical mass for
the development of  CNO flashes ($\sim 0.2M_{\sun}$).  This has been 
motivated by the fact that at least two of
the three ELM pulsating white dwarfs reported by Hermes et al. (2013) have stellar
masses near this threshold value.  Specifically, we
have  compared  the period  spacings  of  a  model with  $M_*=0.1762
M_{\sun}$, corresponding  to the most massive ELM  sequence which does
not experience CNO flashes  on the cooling branch, with the
period spacings of a  model with $M_*=0.1805 M_{\sun}$, corresponding
to the  lowest mass  He core  white dwarf  sequence that  experiences CNO
flashes. Despite the small difference in their stellar masses, we  have  
found  that  these   models  have
appreciably different internal chemical structures,  which impact quite
differently on the  mode trapping  properties of the models for  periods
shorter  than  about  $2500$  s.   So,  we  can  envisage  an
asteroseismic  diagnostic   tool  for  distinguishing   stars  that  have
undergone  CNO flashes in their  early cooling  phase from  those that
have not, provided that enough consecutive pulsation periods of $g$-modes 
with low- and intermediate radial orders and the same harmonic degree were 
detected in pulsating ELM white dwarfs.

Finally, although the cooling age and mass of observed  He-core white
dwarfs can be directly inferred from the information provided
in this paper, the complete set of our  evolutionary sequences  can be 
found at our web site {\tt http://www.fcaglp.unlp.edu.ar/evolgroup}.


\begin{acknowledgements}
We warmly acknowledge the comments and suggestions of our referee, which strongly improved 
the original version of this paper. This  research   was  supported   by  PIP
112-200801-00940 from  CONICET and by AGENCIA through the Programa de
Modernizaci\'on Tecnol\'ogica BID 1728/OC-AR.
\end{acknowledgements}



\begin{thebibliography}{} 
 \bibitem[]{} Althaus, L. G., Serenelli, A. M., \& Benvenuto, O. G., 
2001, MNRAS, 323, 471
\bibitem[Althaus  et al. (2010a)]{2010AARv..18..471A}  Althaus, L.~G.,
  C{\'o}rsico, A.~H.,  Isern, J., \&  Garc{\'{\i}}a--Berro, E.\ 2010a,
  \aapr, 18, 471
\bibitem[]{}  Althaus, L.~G., Serenelli,  A.  M.,  C{\'o}rsico, A.~H.,
  Montgomery, M. H., 2003, \aap, 404, 593
\bibitem[]{}  Althaus,  L.~G.,  Garc{\'{\i}}a--Berro, E.,  Isern,  J.,
  C{\'o}rsico, A.~H., \& Miller Bertolami, M. M., 2012, \aap, 537, 33
\bibitem[Althaus  et al. (2007)]{2007A&A...465..249A}  Althaus, L.~G.,
  Garc{\'{\i}}a--Berro,   E.,  Isern,   J.,  C{\'o}rsico,   A.~H.,  \&
  Rohrmann, R.~D.\ 2007, \aap, 465, 249
\bibitem[]{}  Althaus,  L.~G.,  Serenelli,  A.   M.,  Panei,  J.   A.,
  C\'orsico, A.   H., Garc{\'{\i}}a--Berro, E., \&  Sc\'occola, C. G.,
  2005, \aap, 435, 631
\bibitem[Althaus  et al. (2010)]{2010ApJ...717..897A}  Althaus, L.~G.,
  C{\'o}rsico,  A.~H., Bischoff-Kim,  A., Romero,  A. D.,  Renedo, I.,
  Garc\'ia-Berro, E.,  \& Miller Bertolami,  M. M., 2010b,  \apj, 717,
  897
\bibitem{} Angulo, C., et al. 1999, Nuclear Physics A, 656, 3
\bibitem{} Benvenuto, O. G., \& De Vito, M. A., 2005, \mnras, 362, 891
\bibitem{} Bono, G., Salaris, M., \& Gilmozzi, R., 2013, \aap, 549, A102
\bibitem{} Bradley, P. A. 1996, ApJ, 468, 350
\bibitem[]{} Brown, W. R., Kilic, M., Allende Prieto, C., \& Kenyon S.,
2010, \apj, 723, 1072
\bibitem[]{} Brown, W. R., Kilic, M., Allende Prieto, C., \& Kenyon S.,
2012, \apj, 744, 142
\bibitem[Cassisi  et  al.   (2007)]{2007ApJ...661.1094C} Cassisi,  S.,
  Potekhin,  A.~Y.,   Pietrinferni,  A.,  Catelan,   M.,  \&  Salaris,
  M.\ 2007, \apj, 661, 1094
\bibitem[]{} Chen, X., \& Han, Z., 2002, MNRAS, 335, 948
\bibitem{} C\'orsico, A. H., \& Althaus, L. G. 2006, A\&A, 454, 863
\bibitem{} C\'orsico, A. H., Althaus, L. G., Miller Bertolami, M. M., 
et al. 2012a, MNRAS, 424, 2792 
\bibitem{} C\'orsico, A. H., Althaus, L. G., Romero, A. D., et al. 2012b, 
JCAP, 12, 10 
\bibitem[]{} C{\'o}rsico, A. H., Romero, A. D., Althaus, L. G., \& 
Hermes, J. J., 2012, A\&A, 547, A96
\bibitem{} C\'orsico, A. H., Althaus, L. G., Benvenuto, O. G., \& 
Serenelli, A. M. 2002, A\&A, 387, 531
\bibitem[]{} Driebe, T.,  {Sch\"onberner}, D.,  {Bl\"ocker}, T., \& {Herwig}, F., 1998,
\aap, 339, 123
\bibitem[]{} Ergma, E., Sarna, M. J., \& Gerskevits-Antipova, J., 2001, \mnras, 321, 71 
\bibitem[]{} Ferguson, J. W., Alexander, D. R., Allard, F., Barman T.,
  Bodnarik, J.  G., Hauschildt,  P.  H., Heffner-Wong, A., \& Tamanai,
  A., 2005, ApJ, 623, 585
\bibitem[]{} Fontaine, G., \&  Brassard, P., 2008, PASP, 120, 1043
\bibitem[Garcia-Berro  et  al.  (1995)]{1995MNRAS.277..801G}  Garc\'\i
  a--Berro,  E., Hernanz,  M., Isern,  J., \&  Mochkovitch,  R.\ 1995,
  \mnras, 277, 801
\bibitem[Garc{\'{\i}}a-Berro   et   al.   (2010)]{2010Natur.465..194G}
  Garc{\'{\i}}a--Berro, E., Torres, S.,  Althaus, L.~G., et al.\ 2010,
  \nat, 465, 194
\bibitem[Garc{\'{\i}}a-Berro   et   al.   (2011)]{2011JCAP...05..021G}
  Garc{\'{\i}}a--Berro,   E.,  Lor{\'e}n--Aguilar,  P.,   Torres,  S.,
  Althaus, L.~G., \& Isern, J.\ 2011, \jcap, 5, 21
\bibitem[]{} Gautschy, A., 2013, arXiv:1303.6652
\bibitem[Hansen et  al. (2007)]{2007ApJ...671..380H} Hansen, B.~M.~S.,
  Anderson, J., Brewer, J., et al.\ 2007, \apj, 671, 380
\bibitem[]{} Hermes, J. J., Montgomery, M. H., Winget, D. E., et al., 2012, 
ApJL, 750, L28
\bibitem[]{} Hermes, J. J., Montgomery, M. H., Winget, D. E., et al., 2013, 
\apj, 765, 102
\bibitem[Iglesias  \&  Rogers  (1996)]{1996ApJ...464..943I}  Iglesias,
  C.~A., \& Rogers, F.~J.\ 1996, \apj, 464, 943
\bibitem[Isern et al. (1992)]{1992ApJ...392L..23I} Isern, J., Hernanz,
  M., \& Garc\'\i a--Berro, E.\ 1992, \apjl, 392, L23
\bibitem[Isern   et  al.    (2008)]{2008ApJ...682L.109I}   Isern,  J.,
  Garc{\'{\i}}a--Berro,  E.,  Torres, S.,  \&  Catal{\'a}n, S.\  2008,
  \apjl, 682, L109
\bibitem[]{} Kepler, S. O., Kleinman, S. J., Nitta, A., Koester, D., 
Castanheira, B. G., Giovannini, O., Costa, A. F. M., \& Althaus, L. G, 2007, \mnras,
375, 1315 
\bibitem[]{} {Koester}, D., {Voss}, B.,  {Napiwotzki}, R., {Christlieb}, N., 
	{Homeier}, D., {Lisker}, T., {Reimers}, D., \&  {Heber}, U., 2009,
         \aap, 505, 441
\bibitem[]{} Magni, G. \& Mazzitelli, I., 1979, \aap, 72, 134
\bibitem[]{} Maxted, P.~F.~L., Anderson, D. R., Burleigh, M. R., et al., 2011, \mnras, 418, 1156
\bibitem[]{} Miller Bertolami, M. M., Althaus, L. G., Olano, C., \& Jim\'enez, 
N., 2011a, MNRAS, 415, 1396
\bibitem[]{} Miller Bertolami, M. M., Althaus, L. G., Unglaub, K., \& Weiss, 
A., 2008, A\&A, 491, 253
\bibitem[]{} Miller Bertolami, M. M., Rohrmann, R. D., Granada, A., \& 
Althaus, L. G., 2011b, ApJ, 743, L33
\bibitem[]{} Muslimov, A. G., \& Sarna, M. J. ,1993, MNRAS, 262, 164
\bibitem[]{} Nelson, L. A., Dubeau, E., \& MacCannell, K. A., 2004, \apj, 
616, 1124
\bibitem[]{} Panei, J. A., Althaus, L. G., Chen, X., \& Han, Z., \mnras, 
2007, 382, 779
\bibitem[Renedo  et   al.   (2010)]{2010ApJ...717..183R}  Renedo,  I.,
  Althaus, L.~G.,  Miller Bertolami, M.~M.,  et al.\ 2010,  \apj, 717,
  183
\bibitem[{{Rohrmann} {et~al.}(2012){Rohrmann}, {Althaus},
  {Garc{\'{\i}}a-Berro}, {C{\'o}rsico}, \& {Miller Bertolami}}]{rohrmann}
{Rohrmann}, R.~D., {Althaus}, L.~G., {Garc{\'{\i}}a-Berro}, E., {C{\'o}rsico},
  A.~H., \& {Miller Bertolami}, M.~M. 2012, \aap, 546, A119

\bibitem{} Salaris, M., Althaus, L. G., \& Garc{\'{\i}}a--Berro, E., 2013,
\aap, in press
\bibitem{}  Sarna  M.  J.,  Ergma E., Antipova  J., 2000, \mnras, 316, 84
\bibitem[]{}  Serenelli, A.  M., Althaus,  L.  G.,  Rohrmann,  R.  D.,
  Benvenuto, O. G., 2002, MNRAS, 337, 1091
\bibitem[]{} Silvotti, R., Ostensen, R. H., Bloemen, S., et al. 2012, MNRAS, 424, 1752
\bibitem[]{} Steinfadt, J. D. R., Bildsten, L., \& Arras, P., 2010, \apj,718, 441
\bibitem{} Unno, W., Osaki, Y., Ando, H., Saio, H., \& Shibahashi, H. 
1989, {\it Nonradial Oscillations of Stars} 
(University of Tokyo Press), 2nd. edn.
\bibitem[von   Hippel  \&  Gilmore   (2000)]{2000AJ....120.1384V}  von
  Hippel, T., \& Gilmore, G.\ 2000, \aj, 120, 1384
\bibitem[]{} Wachlin, F. C.,  Miller Bertolami, M. M., \& Althaus, L. G., 
2011, A\&A, 533, A139
\bibitem[]{} Weiss, A. \& Ferguson, J. 2009, A\&A, 508, 1343
\bibitem[]{} Winget, D. E., \& Kepler, S. O., 2008 ARAA, 46, 157
\bibitem[Winget  et al.   (2004)]{} Winget,  D.~E., Sullivan,  D.  J.,
  Metcalfe, T. S., Kawaler, S.  D., \& Montgomery, M. H., 2004, \apjl,
  602, L109
\bibitem[Winget  et al.   (2009)]{2009ApJ...693L...6W}  Winget, D.~E.,
  Kepler, S.~O., Campos, F., et al.\ 2009, \apjl, 693, L6
\end{thebibliography}
\end{document}